\documentclass[aps,prb,amsmath,amssymb,floatfix,twocolumn,amsmath,superscriptaddress,
twocolumn,nofootinbib]{revtex4-2}
\usepackage{color,epsfig}
\usepackage{bm}
\usepackage{amsmath,amsfonts,amssymb,bm}
\usepackage{hyperref}
\usepackage{subfigure}
\usepackage{feynmf}

\newcommand{\bea}{\begin{eqnarray}}
\newcommand{\eea}{\end{eqnarray}}
\newcommand{\be}{\begin{equation}}
\newcommand{\ee}{\end{equation}}

\renewcommand\Re{\text{Re}}
\renewcommand\Im{\text{Im}}

\begin{document}

\title{Monopole versus spherical harmonic superconductors: Topological repulsion, coexistence and stability}
\author{Enrique Mu\~{n}oz}
\affiliation{Facultad de F\'isica, Pontificia Universidad Cat\'olica de Chile, Vicu\~{n}a Mackenna 4860, Santiago, Chile}
\affiliation{Research Center for Nanotechnology and Advanced Materials CIEN-UC, Pontificia Universidad Cat\'olica de Chile, Vicu\~{n}a Mackenna 4860, Santiago, Chile}
\author{Rodrigo Soto-Garrido}
\affiliation{Facultad de F\'isica, Pontificia Universidad Cat\'olica de Chile, Vicu\~{n}a Mackenna 4860, Santiago, Chile}
\author{Vladimir Juri\v ci\' c}
\affiliation{Nordita, KTH Royal Institute of Technology and Stockholm University, Roslagstullsbacken 23, 10691 Stockholm, Sweden}

\date{\today}

\begin{abstract}
The monopole harmonic superconductor (SC), proposed in  doped Weyl semimetals as a pairing between the Fermi surfaces enclosing the Weyl points, is rather unusual, as it features the monopole charge inherited from the parent  metallic phase. However, this state can compete with more conventional spherical harmonic pairings, such as an $s$-wave. We here demonstrate, within the framework of the weak coupling mean-field BCS theory, that the monopole and a conventional spherical harmonic SC quite generically coexist, while the repulsion can take place when the absolute value of the monopole charge matches the angular momentum quantum number of the spherical harmonic. As we show, this feature is a direct consequence of the topological nature of the monopole SC, and we dub it \emph{topological repulsion}. We illustrate the above principle with the example of  the conventional  $s-$ and $(p_x\pm ip_y)-$wave pairings  competing with the monopole SC $Y_{-1,1,0}(\theta,\phi)$,  which coexist in a finite region of the parameter space, and repel, respectively. Furthermore, the s-wave pairing is more stable both when the chemical potentials at the nodes are unequal, and in the presence of point-like charged impurities. Since the phase transition is discontinuous, close to the phase boundary,  we predict that the Majorana surface modes at the interfaces between domains featuring the monopole and the trivial phases, such as an $s-$wave, will be the experimental signature of the monopole SC.
\end{abstract}

\maketitle


\section{Introduction}

Topological semimetals feature the nodal points in the Brillouin zone  where the conduction and valence bands touch, yielding a rather rich landscape of emergent low-energy quasiparticles ~\cite{volovik-book,balatsky-review2014,Yang-NatComm2014,Chiu-RMP2016,Bradlyn-Science2016,Wieder-PRL2016}.
In particular, the exotic electronic properties in Weyl semimetals (WSMs), such as Fermi arc surface states and anomalous magnetotransport, arise from the two topological nodal points in the Brillouin zone featuring pseudorelativistic Weyl fermions~\cite{Burkov-NatMat2016,Jia-NatMat2016,Hasan-ARCP2017,Armitage-RMP2018,bernevig-JPSJ2018}, which were experimentally observed in mostly binary compounds, such as TaAs and NbP~\cite{Xu-Science2015,Lv-PRX2015,Zhang-PRB2017,Xu-NatPhys2015}. These Weyl points are the source and the sink of the Abelian Berry curvature, yielding the monopole charge ${\mathcal C}=\pm1$, the topological invariant characterizing these semimetals.  Weyl metals can also  represent a platform for the realization of yet different states of matter. For instance,   they can host an interaction-driven fully gapped axionic insulator~\cite{Wang-Zhang-PRB2013,Roy-Sau-PRB2015,You-PRB2016,Roy-PRB2017}, which was recently experimentally observed~\cite{Gooth-Nature2019}.
On the superconducting side, WSMs can accommodate a plethora of pairing states~\cite{Volovik-jetp87,Murakami-PRL2003,Meng-PRB2012,Cho-PRB2012,Yang-PRL2014,Bednik-PRB2015,Schnyder-JPCM2015}. The monopole superconductor (SC),  recently proposed as a pairing state between the two Fermi surfaces (FSs) enclosing the Weyl points in a doped WSM~\cite{Li-Hadane-PRL2018}, is an exciting possibility because it hosts vortices inherited from the underlying WSM state, but its physical consequences have been only touched upon so far~\cite{Sun-arxiv2019}.

\begin{figure*}[t!]
    \centering
    \includegraphics[width=0.46\textwidth]{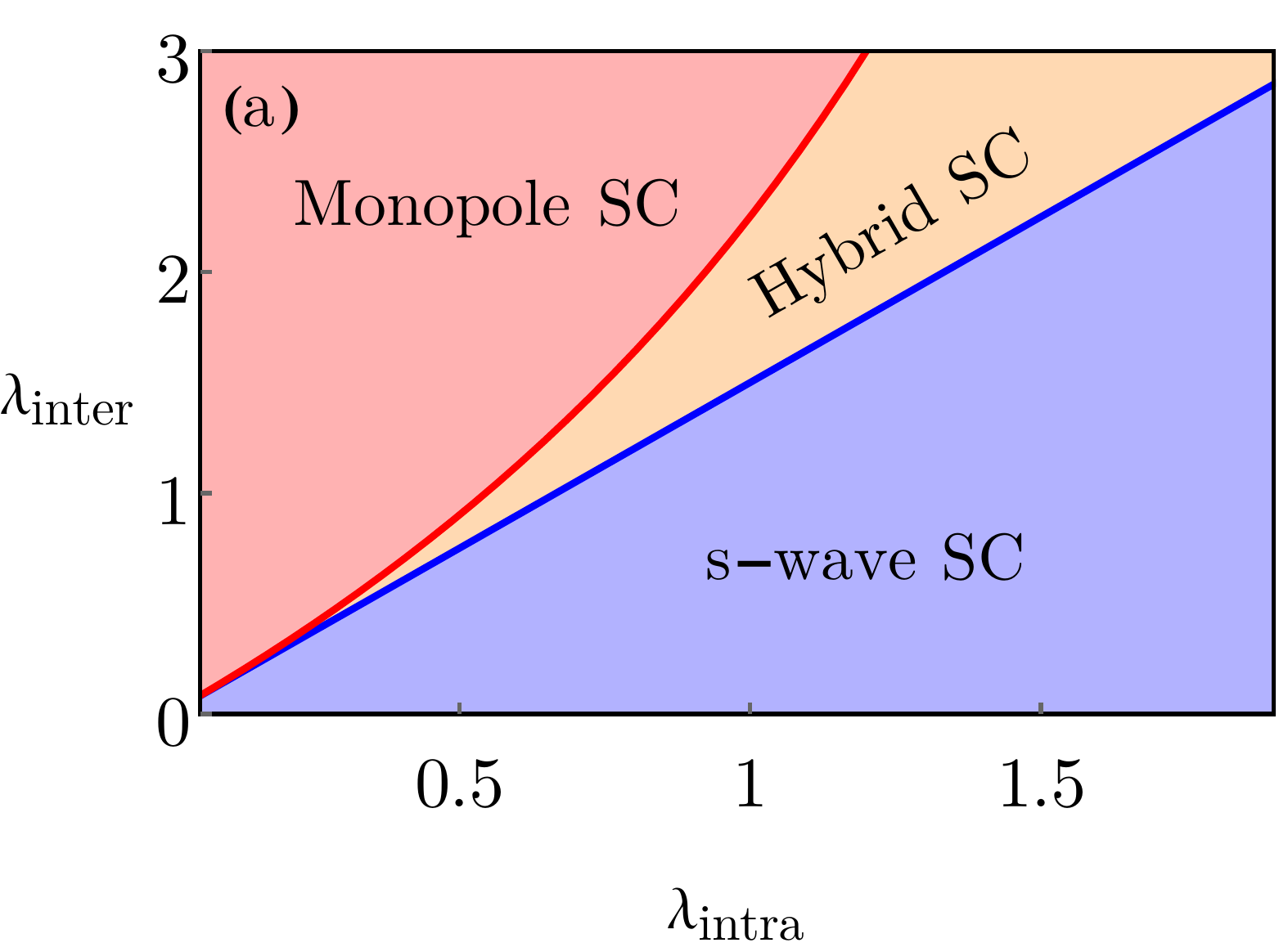}
    \hspace{.1cm}
    \includegraphics[width=0.46\textwidth]{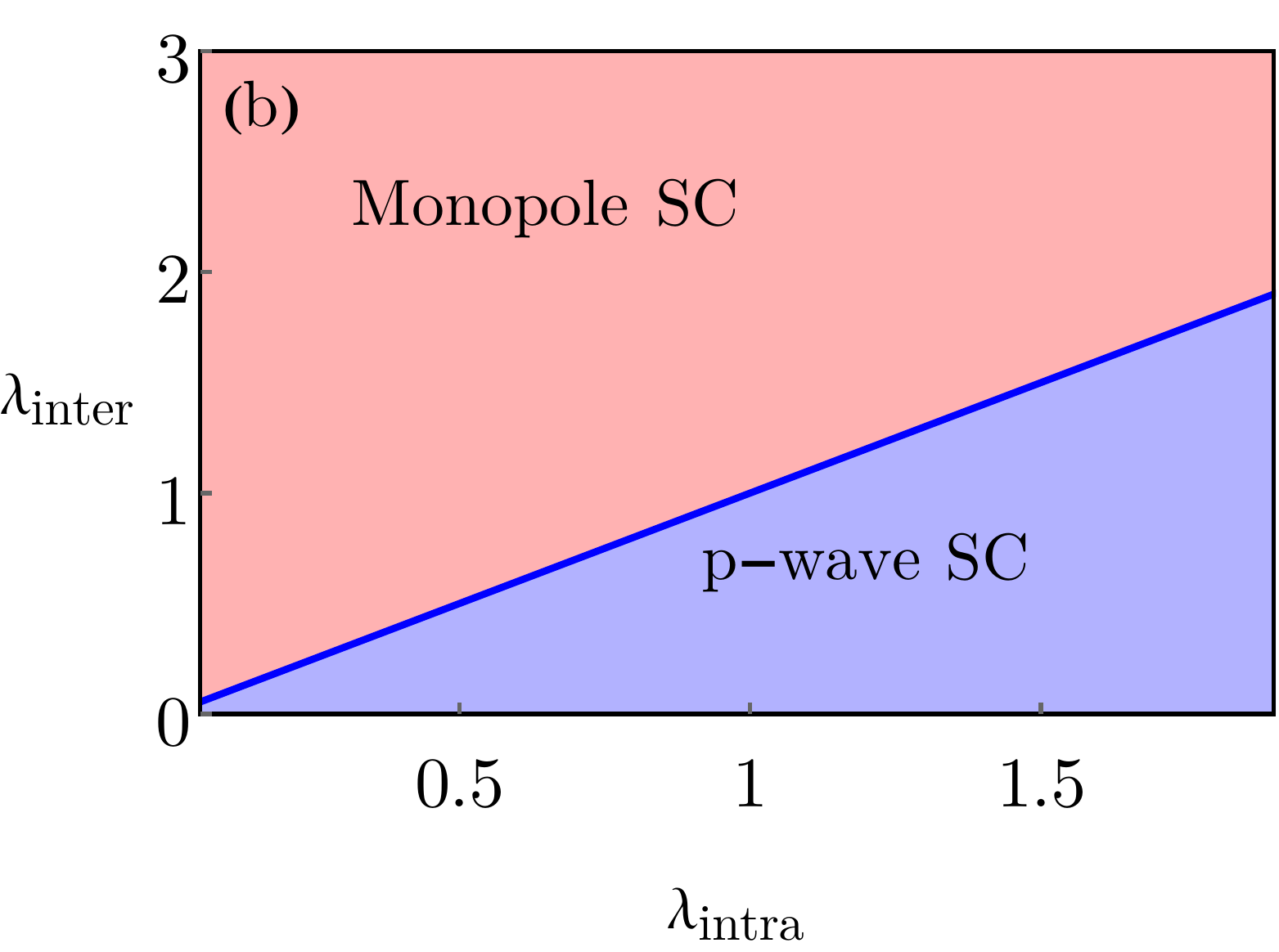}
    \caption{Zero-temperature phase diagram of the Weyl superconductor (SC) with the inter- and intra-Fermi surface pairings tuned by the couplings $\lambda_{inter}$ and $\lambda_{intra}$, respectively. (a)  Monopole versus the $s$-wave pairing. The two phases coexist in a finite region of the phase diagram (hybrid SC). The phase boundaries between mixed and  pure phases are given by Eq.~\eqref{eq:phaseboundary-swave} (blue solid line) and Eq.~\eqref{eq:phaseboundary-monopole} (red solid line). (b) Monopole versus the $p_\pm$-wave SC. Topological repulsion is operative thus the phases may coexist only at the phase boundary where the two couplings are equal, see the discussion after Eq.~\eqref{eq:BCS-equal} and Sec.~\ref{subsection:monopolevspwave}. The phase transitions between different phases are discontinuous. The effective couplings $\lambda_{inter}$ and $\lambda_{intra}$ are defined in Eq.~\eqref{eq:effective-couplings}.  }
    \label{fig:phasediagram}
\end{figure*}

 An urgent issue in this respect is the competition of the monopole pairing, characterized by the monopole harmonic functions $Y_{q,j,m}(\theta,\phi)$ with more conventional spherical harmonic states $Y_{j,m}(\theta,\phi)$, as well as  its stability in the presence of impurities. We here demonstrate, within the framework of the mean-field BCS theory, that the monopole SC and a conventional spherical harmonic phase quite generically can coexist with one another, while the repulsion takes place when the $\theta-$dependent form factors  of the monopole harmonic $Y_{|q|,j,|m|}(\theta,\phi)$ and the spherical harmonic $Y_{q=0,j,m}(\theta,\phi)\equiv Y_{j,m}$ are proportional. As we show, this manifestly gauge-independent  feature, is a direct consequence of the topological nature of the monopole superconductor, and we dub it \emph{topological repulsion}. In particular, this mechanism  implies that a monopole harmonic with the charge $q>0$, $Y_{-q,q,0}(\theta,\phi)$ and  the spherical harmonic $Y_{q,q}(\theta,\phi)$ always repel, with the coexistence possible only at the phase boundary. We illustrate the above principle by showing that the conventional  $s-$wave and a monopole superconductor $Y_{-1,1,0}$  can coexist in a finite region of the parameter space, as shown in Fig~\ref{fig:phasediagram}(a). On the other hand,  this monopole pairing repels the $p_x\pm i p_y$ superconducting states, as displayed in Fig.~\ref{fig:phasediagram}(b).  Furthermore, the s-wave pairing is more stable both when the chemical potentials at the nodes are unequal (Fig.~\ref{fig:deltamu}) and in the presence of point-like charged impurities, see Fig.~\ref{fig:taus}. Since the phase transition is discontinuous, close to the phase boundary,  we predict that the Majorana surface modes at the interfaces between domains featuring the monopole and the trivial phases, such as an $s-$wave, will be the experimental signature of the monopole superconductor in the system.

 The rest of the paper is organized as follows. In Sec.~\ref{sec:Model}, we present the details of the continuum model for the Weyl semimetal. Section~\ref{sec:cleanlimit} is devoted to the mean-field analysis of the competition and coexistence  between the monopole and spherical harmonic pairings corroborated by  a gauge invariance argument for the topological repulsion. In Sec.~\ref{sec:scattering}, we analyze the effects of the impurity scattering on the competing $s-$wave and monopole SCs. Finally, in Sec.~\ref{sec:discussion}, we discuss our results and we present necessary technical details in the appendices.


\section{Model}
\label{sec:Model}

We start by considering the model describing the mean-field Cooper pairing  between the Weyl quasiparticles living at the FSs enclosing the two nodal points with opposite monopole charges ${\cal C}_\pm=\pm 1$
\begin{eqnarray}
\hat{H} = \hat{H}_{Weyl} + \hat{H}_{\Delta}.
\label{eq_1}
\end{eqnarray}
The continuum Hamiltonian corresponding to the time-reversal symmetry breaking WSM with the two nodal points is $\hat{H}_{Weyl} =\sum_{\zeta=\pm,\mathbf{q}}\hat{c}_{\zeta{\bf K}_0+\mathbf{q}}^{\dagger}\hat{h}_\zeta({\bf q}) \hat{c}_{\zeta{\bf K}_0+\mathbf{q}}$, where
\begin{equation}
\label{eq:continuum-Weyl}
\hat{h}_\zeta({\bf q}) = v_F(\sigma_x q_x+\sigma_y q_y+\zeta \sigma_z q_z)-\mu,
\end{equation}
and the chemical potential $\mu>0$.
This Hamiltonian is obtained after expanding the corresponding lattice model  about the two Weyl nodes along the $k_z$-direction located at $\zeta\mathbf{K}_0 = \left(0,0,\zeta K_0 \right)$, (see Appendix~\ref{app:Model}).
We here consider only isotropic nodes with Fermi velocity $v_F=1$, and fix the position of the nodes at  $K_0 = \pi/2a$, with the lattice constant $a=1$, and also $\hbar=k_B=1$ hereafter.

We here adopt a phenomenological model in which we assume both inter-FS and intra-FS pairing interactions without delving into microscopic details yielding such interactions. This allows us to address the universal aspects of the competition and coexistence between the spherical and monopole harmonics in a Weyl superconductor, which is the main purpose of this paper.
The Hamiltonian for an inter-FS s-wave spin-singlet pairing is
\begin{equation}
\label{eq:pairing-interband-full}
\hat{H}_{\Delta} = \sum_{\mathbf{q}}\hat{c}_{{\bf K}_0+\mathbf{q}}^{\dagger}[{\bar\Delta}_0 i\hat{\sigma}_y] \hat{c}_{-{\bf K}_0-\mathbf{q}}^{\dagger} + H.c.,
\end{equation}
with ${\bar\Delta}_0$ being the order parameter. This is possibly  the simplest pairing between  the Fermi surfaces $FS_{\pm}$ enclosing  the two nodal points at $\zeta{\bf K}_0$ and involves the two Weyl quasiparticles with momenta ${\bf K}_0+{\bf q}$ and $-{\bf K}_0-{\bf q}$, where  $\pm{\bf q}$ lives on the sphere $S_\pm$ obtained after shifting $FS_\pm$ by $\mp{\bf K}_0$ toward the origin. Crucially, the Cooper pair wavefunction acquires the total Berry flux $4\pi$ inherited from the parent chiral Weyl fermions~\cite{Murakami-PRL2003}. Consequently, its projection onto the sphere $S_+$ ($S_-$)  features at least one vortex with the unit ($2\pi$) vorticity, and the corresponding projected pairing is proportional to a monopole harmonic  function $Y_{q,l,m}(\theta,\phi)$, with $4\pi q$ counting the total Berry flux of the SC state~\cite{Li-Hadane-PRL2018}.

More formally, the band basis on the Fermi surfaces $FS_{\pm}$ is $\hat{\alpha}^{\dagger}_{\pm}(\pm\mathbf{q}) = \sum_{\sigma=\uparrow,\downarrow}\xi_{\pm,\sigma}(\pm\mathbf{q})\hat{c}^{\dagger}_{\pm\mathbf{K}_0\pm\mathbf{q},\sigma}$, with the spinors  $\xi_{\pm,\uparrow}(\pm\mathbf{q}) = \left(u_{\mathbf{q}}, v_{\mathbf{q}} \right)^T$,
chosen so that the Dirac string pierces the sphere at the south pole (spherical polar angle $\theta_{\bf q}=\pi$),
since  $u_{\mathbf{q}} = \cos\left(\theta_{\mathbf{q}}/2 \right)$ and $v_{\mathbf{q}} = \sin\left(\theta_{\mathbf{q}}/2 \right) e^{i\phi_{\mathbf{q}}}$, and $\phi_{\mathbf{q}}$ is the azimuthal angle.
After projecting the pairing Hamiltonian in Eq.~\eqref{eq:pairing-interband-full} onto the FS$_\pm$ [$\pm{\bf q}\in FS_\pm$] in the weak coupling (BCS) regime $|{\bar\Delta}_0|\ll |\mu|$, we obtain
\begin{eqnarray}
\hat{\tilde{H}}_{\Delta} = \sum_{\mathbf{q}}\hat{\alpha}_{-}^{\dagger}(\mathbf{q})\tilde{\Delta}(\mathbf{q})\hat{\alpha}^{\dagger}_{+}(-\mathbf{q}) + H.c.
\label{eq_pairing}
\end{eqnarray}
with the gap function  $\tilde{\Delta}(\mathbf{q}) = -2{\bar\Delta}_0 u_{\mathbf{q}}^* v_{\mathbf{q}}^* = -\bar{\Delta}_0 \sin\theta_{\mathbf{q}} e^{-i\phi_{\mathbf{q}}}=-{\bar \Delta}_0\sqrt{\frac{4\pi}{3}}Y_{-1,1,0}(\theta_{\bf q},\phi_{\bf q})$, where $Y_{q,l,m}(\theta,\phi)$ is the standard monopole harmonic function~\cite{CNY-NPB1976,Haldane-PRL1983}.
Notice that for the  monopole pairing in Eq.~\eqref{eq_pairing},  $2q=2\mathcal{C}_-=-2$, since $\mathcal{C}_\pm\rightarrow \mathcal{C}_\mp$ under ${\bf q}\rightarrow -{\bf q}$.

In a WSM prone to a superconducting instability, a more conventional intra-FS spin-singlet pairing, which necessarily occurs at a finite momentum $2{\bf K}_0$, is also possible, and competes with the monopole SC. Furthermore, the inversion symmetry in  Weyl materials may be broken, so to account for this effect, we consider
slightly different chemical potentials at  the two nodes, $\mu_{-}$
and $\mu_{+}$, with $|\delta\mu| = |\mu_{+} - \mu_{-}| \ll \bar{\mu}$,
where $\bar{\mu} = \left(\mu_{+} + \mu_{-} \right)/2$ is the average chemical potential. The mean-field Bogoliubov-de Gennes Hamiltonian that includes both pairing instabilities takes the form
\begin{eqnarray}\label{eq:HBdG}
\hat{H} = \sum_{\mathbf{q}}\Psi_{\mathbf{q}}^{\dagger}\hat{H}_{BdG}(\mathbf{q})\Psi_{\mathbf{q}},
\end{eqnarray}
with
\begin{equation}
\hat{H}_{BdG}(\mathbf{q}) = \left[\begin{array}{cccc}\xi_{\bf q}^{-} & \Delta_0 & 0 & \tilde{\Delta}_{\mathbf{q}}\\
\Delta_{0}^{*} & -\xi_{\bf q}^{-} & \tilde{\Delta}_{\mathbf{q}}^{*} & 0\\
0 & \tilde{\Delta}_{\mathbf{q}} & \xi_{\bf q}^{+} & \Delta_0\\
\tilde{\Delta}_{\mathbf{q}}^{*} & 0 & \Delta_0^{*} & -\xi_{\bf q}^{+}\end{array}\right],
\label{eq_H1}
\end{equation}
and the Nambu basis is $\Psi_{\mathbf{q}}^{\dagger} = \left[{\hat\alpha}_{-}^{\dagger}(\mathbf{q}),{\hat\alpha}_{-}(-\mathbf{q}),{\hat\alpha}_{+}^{\dagger}(\mathbf{q}),
{\hat\alpha}_{+}(-\mathbf{q})\right]$, while $\xi_{\bf q}^{\pm} =  v_F |\mathbf{q}| - \mu_{\pm}$. In this basis, we can treat both s-wave and p-wave intra-FS pairings as long as they \emph{separately} compete with the monopole SC. Otherwise, the spin index would have to be explicitly restored and the basis would therefore be doubled.

It is convenient to express the above Hamiltonian by using the $SU(2)\otimes SU(2)$ matrices  $\left\{\hat{\tau}_{\alpha}\otimes \hat{\eta }_{\beta}\right\}$, with $\hat{\tau}_{\alpha}$ representing the $SU(2)$ nodal basis, while $\hat{\eta}_{\beta}$ is the particle-hole basis. Here, $\left\{\hat{\tau}_{\alpha},\hat{\eta }_{\beta}\right\}$ are the Pauli matrices, while $\hat{\tau}_0,\hat{\eta}_0$ are the $2\times2$ unity matrices.   Therefore, the Hamiltonian in Eq.~\eqref{eq_H1} in this representation reads
\begin{align}
 &\hat{H}_{BdG}(\mathbf{q}) = \bar{\xi}_{ q}\hat{\tau}_0\otimes\hat{\eta}_3 + \frac{\delta{\mu}}{2}\hat{\tau}_3\otimes\hat{\eta}_3 + \Re\Delta_0\hat{\tau}_0\otimes\hat{\eta}_1 \nonumber\\
 &- \Im\Delta_0 \hat{\tau}_0\otimes\hat{\eta}_2+\Re \tilde{\Delta}_{\mathbf{q}}\hat{\tau}_1\otimes\hat{\eta}_1
 - \Im \tilde{\Delta}_{\mathbf{q}}\hat{\tau}_1\otimes\hat{\eta}_2,
 \label{eq:HBdG2}
 \end{align}
 where we defined $\bar{\xi}_{\bf q} = v_F|\mathbf{q}| - \bar{\mu}$.

\section{BCS mean-field gap equations: Clean limit}
\label{sec:cleanlimit}

The mean-field gap equations for the two competing superconducting orderings are obtained from the finite-temperature Green's function for the effective Bogoliubov-de Gennes Hamiltonian in Eq.~(\ref{eq:HBdG}), which  in terms of the valley sub-blocks reads as (see Appendix~\ref{app:fieldtheory})
\begin{equation}\label{eq:greens-function}
\hat{\mathcal{G}}_0(\omega_n,\mathbf{q}) = \left[-i\omega_n + \hat{H}_{BdG}(\mathbf{q}) \right]^{-1}=\left[\begin{array}{cc}
\hat{G}_{0}^{--}&\hat{G}_0^{-+}\\\hat{G}_0^{+-}&\hat{G}_0^{++}\end{array}\right],
\end{equation}
where $\hat{G}_0^{\rho\zeta}$, $\rho,\zeta=\pm$, are the $2\times2$ submatrices, and $\omega_n=(2n+1)\pi T$ is the fermionic Matsubara frequency at temperature $T$. The gap equations for the
conventional intra-Fermi surface pairing  and the monopole SC then can be compactly written as
\begin{equation}\label{eq:gap-compact}
\Delta_{\eta}({\bf q}) = -T\sum_{\mathbf{q},\omega_n}V_\eta(\mathbf{q},{\bf q}')
\langle\hat{\alpha}_{-}(\mathbf{q}')\hat{\alpha}_{\zeta}(-\mathbf{q}')\rangle,
\end{equation}
where $\zeta=-$ ($\zeta=+$) for  $\eta=intra$ ($\eta=inter$) corresponding to the spherical (monopole) harmonic pairing, and $V_\eta$ are the pairing potentials. In terms of the Green's function in Eq.~\eqref{eq:greens-function},  $\langle\hat{\alpha}_{-}(\mathbf{q})\hat{\alpha}_\zeta(-{\bf q})\rangle= \left[\hat{G}_0^{-\zeta} \right]_{21}$. The explicit form of the gap equations is given by Eqs.~\eqref{eq:BCS_intra} and~\eqref{eq:BCS_inter}, from which we can conclude that when these two superconducting orders compete new instabilities can be generated but in the insulating (particle-hole) channels. More specifically, when the intra-FS pairing is $s$-wave,  the two $p-$wave charge-density wave orders in the $x-$ and $y-$ directions may get generated. This is so when the two superconducting orders coexist, which is indeed possible, unless the two orders exhibit the same $\theta$-dependent form factors, as we  show below. The study of the effects of a generated insulating state on the superconducting instabilities is, however, beyond the scope of the current work.

The pairing potentials for the spherical harmonics and the monopole channels when $\mu_+=\mu_-=\mu$, dictated by the form of the corresponding pairing functions, are in general given by
\begin{eqnarray}
V_{intra}(\mathbf{q},\mathbf{q}') &=& V_0Y_{l,m}(\theta_\mathbf{q},\phi_\mathbf{q})Y_{l,m}^*(\theta_\mathbf{q'},\phi_\mathbf{q'}),\nonumber\\
V_{inter}(\mathbf{q},\mathbf{q}') &=& \tilde{V}_0 Y_{q,j,m}(\theta_\mathbf{q},\phi_\mathbf{q})Y_{q,j,m}^*(\theta_\mathbf{q'},\phi_\mathbf{q'}),
\end{eqnarray}
where $Y_{l,m}(\theta,\phi)=f_l(\theta)e^{im\phi}$ and $Y_{q,j,m}=e^{i(m+q)\phi}g(\theta)$, where   $Y_{q,j,m}(\theta,\phi)$ is the monopole harmonic with the form given by Eq.~\eqref{eq:monopoleSC-general}.

Let us now consider the $T=0$ BCS gap equation for a more general intra-FS pairing as defined above. For convenience, we introduce the notation
\begin{align}
\Delta_{inter}(\mathbf{q}) &= \bar{\Delta}_0\, d_{inter}(\theta)\, e^{i(m+q)\phi}\nonumber\\
\Delta_{intra}(\mathbf{q}) &= \Delta_{lm,0}\, d_{intra}(\theta)\, e^{im'\phi}.
\end{align}
The corresponding generalized  zero-temperature BCS gap equations, derived in Appendix~\ref{app:cleanlimit}, read 
\begin{widetext}
\begin{eqnarray}
\lambda_{\eta}^{-1} &=& \int\frac{d\theta d\phi}{4\pi}
\sin\theta |d_{\eta}(\theta)|^2\left[
2\ln(2\omega_D) - \frac{1}{2}\sum_{s=\pm}\ln(A_{+} + 2 s B_q)
\right],
\label{eq:BCS-general}
\end{eqnarray}
with $\lambda_{\eta}$ ($\eta = inter,intra$) as the effective coupling constants, given by
\begin{equation}
\label{eq:effective-couplings}
\lambda_{inter}=V_0\rho(\mu),\,\,\,
\lambda_{intra}=\frac{{\tilde V}_0\rho(\mu)}{4\pi}\int_0^{2\pi}d\phi\int_0^\pi d\theta \sin\theta |Y_{-1,1,0}(\theta,\phi)|^2,
\end{equation}
see also Eq.~\eqref{eq:coupling} and discussion therein.
Here, we defined the coefficients
\begin{align}
A_{\pm} &=\Delta_{lm,0}^2f_l^{2}(\theta)   \pm \bar{\Delta}_{0}^2g^2(\theta)\nonumber\\
2 B_q &= 2\left[\Re\Delta_{inter}(\mathbf{q})\Re\Delta_{intra}(\mathbf{q})
+ \Im\Delta_{inter}(\mathbf{q})\Im\Delta_{intra}(\mathbf{q})\right]\nonumber\\
&= 2\Delta_{lm,0}\bar{\Delta}_{0}g(\theta) f_l(\theta)\cos\{[m'-(m+q)]\phi\}\equiv B \cos\{[m'-(m+q)]\phi\}.
\label{eq:coefficients}
\end{align}
The integral over the azimuthal angle is calculated as follows after setting $r=m'-(m+q)$:
\begin{align}
&\int_0^{2\pi}d\phi \ln[A_{+} \pm B\cos(r\phi)]
= \sum_{n=1}^{r}\int_{2\pi (n-1)/r}^{2\pi n/r}d\phi
\ln[A_+ \pm B\cos(r\phi)]\nonumber\\
&= \frac{1}{r}\sum_{n=1}^{r}\int_{2\pi (n-1)}^{2\pi n}\ln[A_+ \pm B \cos\phi]
= 2\pi \ln\left[\frac{1}{2}\left(A_+ + \sqrt{A^2_{+} - B^2}\right) \right]\nonumber\\
&= 2\pi \ln\left[\frac{1}{2}(A_+ + |A_{-}|) \right].
\label{eq:mcos}
\end{align}
\end{widetext}

Remarkably, when  $|f_l(\theta)|\sim|g(\theta)|$, i.e. $|f_l(\theta)|$ and $|g(\theta)|$ are the same functions up to a real coefficient,  and $m'\neq m+q$, Eq.(\ref{eq:mcos}) determines the  competition between a monopole SC phase and a conventional spherical harmonic
SC phase since in that case
\begin{align}
A_+ + |A_{-}| = \left\{
\begin{array}{cc}
2\bar{\Delta}_{0}^2g^2(\theta), & \bar{\Delta}_{0} > \Delta_{lm,0}\\
2\Delta_{lm,0}^2 f_l^2(\theta), & \bar{\Delta}_{0} < \Delta_{lm,0}.
\end{array}
\right.
\end{align}

Therefore, we conclude that in this case a sharp boundary exists between the monopole SC phase and the
spherical harmonic $Y_{l,m'}(\theta,\phi)$ pairing, where the stronger coupling dominates.
Analogously, also when  $|f_l(\theta)|\sim|g(\theta)|$ and $m'= m+q$, the gap equations \eqref{eq:BCS-general} imply that the coexistence is possible only when the two couplings are equal. Namely, in that case the two equations reduce to
\begin{equation}
\lambda_{\eta}^{-1} = \int d\theta
\sin\theta |f_l(\theta)|^2
\ln\left[\frac{2\omega_D} {f_l^{2}(\theta)|\Delta_{lm,0}^2 - \bar{\Delta}_{0}^2|}
\right],
\label{eq:BCS-equal}
\end{equation}
for $\eta=intra,inter$. Therefore, provided that $|f_l(\theta)|\sim |g(\theta)|$, $m'>0$, and $m+q>0$, the phases repel each other. As we show below, the last two conditions can be removed because of the gauge freedom in choosing where the Dirac string pierces the Fermi sphere. In other words, a spherical harmonic and the monopole SC repel each other when the corresponding $\theta-$dependent form factors satisfy $|f_l(\theta)|\sim|g(\theta)|$. It turns out that for any monopole harmonic [see Eq.~\eqref{eq:monopoleSC-general}] $Y_{-q,q,0}(\theta,\phi)\sim(\sin\theta)^q e^{iq\phi}$ and  spherical harmonic  $Y_{q,q}(\theta,\phi)\sim(\sin\theta)^q e^{iq\phi}$, the corresponding pairings always repel each other, and we name this mechanism \emph{topological repulsion}.

To show the gauge independence of this principle, we recall that  gauge choices where the Dirac string originating from the monopole goes through the north pole ($\theta=0$) and the south pole ($\theta=\pi$) are related by a coordinate transformation $\theta\rightarrow\pi-\theta$ and $\phi\rightarrow-\phi$. The equivalent gauge classes are given by $Y_{q,j,m}$ and $Y_{-q,j,-m}$, which can be shown as follows.  We start from the form of the monopole harmonic~\cite{Li-Hadane-PRL2018}
\begin{equation}
\label{eq:monopoleSC-general}
Y_{q,j,m}(\theta,\phi)=\sqrt{\frac{2j+1}{4\pi}}e^{i(m+q)\phi}d^j_{m,-q}(\theta),
\end{equation}
where
\begin{align}
d^j_{m,l}(\theta)&=\sqrt{\frac{(j+l)!(j-l)!}{(j+m)!(j-m)!}}\left(\cos\frac{\theta}{2}\right)^{l+m}\left(\sin\frac{\theta}{2}\right)^{l-m}\nonumber\\
&\times P_{j-l}^{l-m,\, l+m}(\cos\theta),
\end{align}
and the function $P_n^{a,b}(x)$ is defined as
\begin{align}
\label{eq:mono-definition}
P_n^{a,b}(x)=&\frac{(-1)^n}{2^n n!}(1-x)^{-a}(1+x)^{-b}\nonumber\\
&\times\frac{d^n}{dx^n}[(1-x)^{a+n}(1+x)^{b+n}].  \end{align}
Then, in the next step one can easily see that
\begin{equation}
d^j_{m,l}(\theta)\rightarrow d^j_{m,l}(\pi-\theta)=(-1)^{j-l}d^j_{-m,l}(\theta),
\end{equation}
and under $\phi\rightarrow-\phi$
\begin{equation}
e^{i(q+m)\phi}\rightarrow e^{i(-q-m)\phi}.
\end{equation}
The last two equations together with the form of the monopole harmonic functions in Eq.~\eqref{eq:mono-definition}, show that the monopole harmonics $Y_{q,j,m}(\theta,\phi)$ and $Y_{-q,j,-m}(\theta,\phi)$ are equivalent with respect to the choice of the direction of the Dirac string. Therefore, the topological repulsion is operative irrespective of the sign of $m+q$ in the monopole harmonic $Y_{q,l,m}(\theta,\phi)$ [see Eq.~\eqref{eq:monopoleSC-general}].

We now discuss special cases when the s-wave and the p-wave pairings compete with the monopole SC $Y_{-1,1,0}(\theta,\phi)$ (the Dirac string pierces the south pole).

\subsection{Monopole versus $s-$wave pairing state}
The above criterion implies that the coexistence of the $s$-wave pairing and the monopole harmonic $Y_{-1,1,0}(\theta,\phi)$ may be possible, as indeed displayed in Fig.~\ref{fig:phasediagram}(a).  This conclusion is based on the solution of the self-consistent gap equations, which read 
\begin{widetext}
\bea
\label{eq:SCBCS}
\lambda_{intra}^{-1} &=& 2\pi\int_0^{\pi}\frac{d\theta}{4\pi}\sin\theta\left(2\ln(2\omega_D) - \ln\left[\frac{1}{2}\left(
\Delta_s^2 + \bar{\Delta}_0^2\sin^2\theta
+ \left|\Delta_s^2 - \bar{\Delta}_0^2\sin^2\theta \right|
\right)
\right] \right)  = 2\ln(2\omega_D) - \frac{1}{2}\mathcal{J}^{(1)}\nonumber\\
\lambda_{inter}^{-1} &=&
2\pi\int_0^{\pi}\frac{d\theta}{4\pi}\sin^3\theta\left(
2\ln(2\omega_D) - \ln\left[\frac{1}{2}\left(
\Delta_s^2 + \bar{\Delta}_0^2\sin^2\theta
+ \left|\Delta_s^2 - \bar{\Delta}_0^2\sin^2\theta \right|
\right)
\right]
\right) = \frac{4}{3}\ln(2\omega_D) - \frac{1}{2}\mathcal{J}^{(3)}.
\label{eq:monop1}
\eea
\end{widetext}
Here, we defined the integrals (for $n=0,1$)
\bea
\mathcal{J}^{(2n+1)} &=& \int_0^{\pi}d\theta\sin^{2n+1}\theta\\
&\times&\ln\left[\frac{1}{2}\left(
\Delta_s^2 + \bar{\Delta}_0^2\sin^2\theta
+ \left|\Delta_s^2 - \bar{\Delta}_0^2\sin^2\theta \right|
\right)
\right].\nonumber
\label{eq:J1}
\eea
To evaluate these angular integrals, one needs to distinguish
two separate regimes in the parameter space in order to handle correctly
the absolute value in the integrand. For $\Delta_s > \bar{\Delta}_0$, it is straightforward to obtain (for $n = 0,1$)
\bea
\mathcal{J}^{(2n+1)} = \frac{2^{n+1}n!}{(2n+1)!!} \ln(\Delta_s^2).
\label{eq:J2}
\eea
On the other hand, for  $\Delta_s<\bar{\Delta}_0$, we define the
angular parameter $\sin\theta_0 = \Delta_s/\bar{\Delta}_0$, which
allows us to calculate the integrals by splitting the domain into the subintervals $\theta \in [0,\theta_0] \cup [\theta_0,\pi-\theta_0] \cup [\pi-\theta_0,\pi]$. The corresponding result is ($\Delta_s<\bar{\Delta}_0$)
\bea
\mathcal{J}^{(1)} &=& 2\ln(\Delta_s^2) + 4\left(\ln\left[\frac{1 + \cos\theta_0}{\sin\theta_0}\right] - \cos\theta_0
\right)\nonumber\\
\mathcal{J}^{(3)} &=& \frac{4}{3}\ln(\Delta_s^2)-\frac{1}{3}
\left(8 - 9\cos\theta_0 + \cos3\theta_0  \right)\nonumber\\
&&+ \frac{1}{9}\left\{
24 \ln\left[ \frac{1+\cos\theta_0}{\sin\theta_0}\right]
+ \cos3\theta_0\left( 1 - 3\ln\sin\theta_0 \right)\right.\nonumber\\
&&\left.+ 3\cos\theta_0(-7 + 9\ln\sin\theta_0)\right\}.
\label{eq:J3}
\eea

To understand the phase diagram displayed in Fig.~\ref{fig:phasediagram}(a), we start from $\lambda_{inter}=0$ and thus $\Delta_0=0$, and find that for any $\lambda_{inter}<3\lambda_s/2$, the system is in the pure $s-$wave state. The line at which the coexistence of these two SC phases sets in is given by the condition $\Delta_s = \bar{\Delta}_0$ ($\theta_0 = \pi/2$), the form is obtained by using Eqs.~(\ref{eq:monop1}) and  Eq.~(\ref{eq:J3}),  and reads as
\bea\label{eq:phaseboundary-swave}
\lambda_{inter}^{-1} = \frac{4}{3}\ln(2\omega_D) - \frac{2}{3}\ln(\Delta_0^2) = \frac{2}{3}\lambda_{intra}^{-1},
\eea
see also additional analysis on the coexistence across this line in Appendix~\ref{app:phaseboundary}.
The transition from the pure $s-$wave to the mixed SC is discontinuous since the solution $\Delta_s\neq0, \Delta_0=0$ is valid up to the phase boundary line (the blue solid line in Fig.~\ref{fig:phasediagram}(a)), and it jumps to $\Delta_s=\Delta_0\neq0$.
As the coupling $\lambda_{intra}$ is further increased across this line,
the coexistence regime persists until the pure monopole SC state is reached, which occurs for $\Delta_s=0$. Using Eq.~\eqref{eq:SCBCS}, we obtain the form of the phase boundary between the hybrid SC and the pure monopole SC [the red solid line in Fig.~\eqref{fig:phasediagram}(b)]:
\begin{equation}\label{eq:phaseboundary-monopole}
\lambda_{inter}^{-1}-\frac{2}{3}\lambda_{intra}^{-1}=-\frac{2}{9}.
\end{equation}
The phase transition from the coexisting region to the pure monopole state is discontinuous, as can be readily shown by inserting the form of the phase boundary given by  Eq.~\eqref{eq:phaseboundary-monopole} into Eqs.~(\ref{eq:monop1}) and~(\ref{eq:J3}) valid for $0\neq\Delta_s<\Delta_0$, which shows that as the boundary is approached from the coexistence region $\Delta_s$ cannot vanish. Finally, we point out that the obtained phase diagram, shown in Fig.~\ref{fig:phasediagram}(a), is qualitatively similar to the one for the mixed $s+id$ superconductor in a two-dimensional Fermi liquid~\cite{Musaelian-PRB1996}.

\begin{figure}[t!]
    \includegraphics[width=0.47\textwidth]{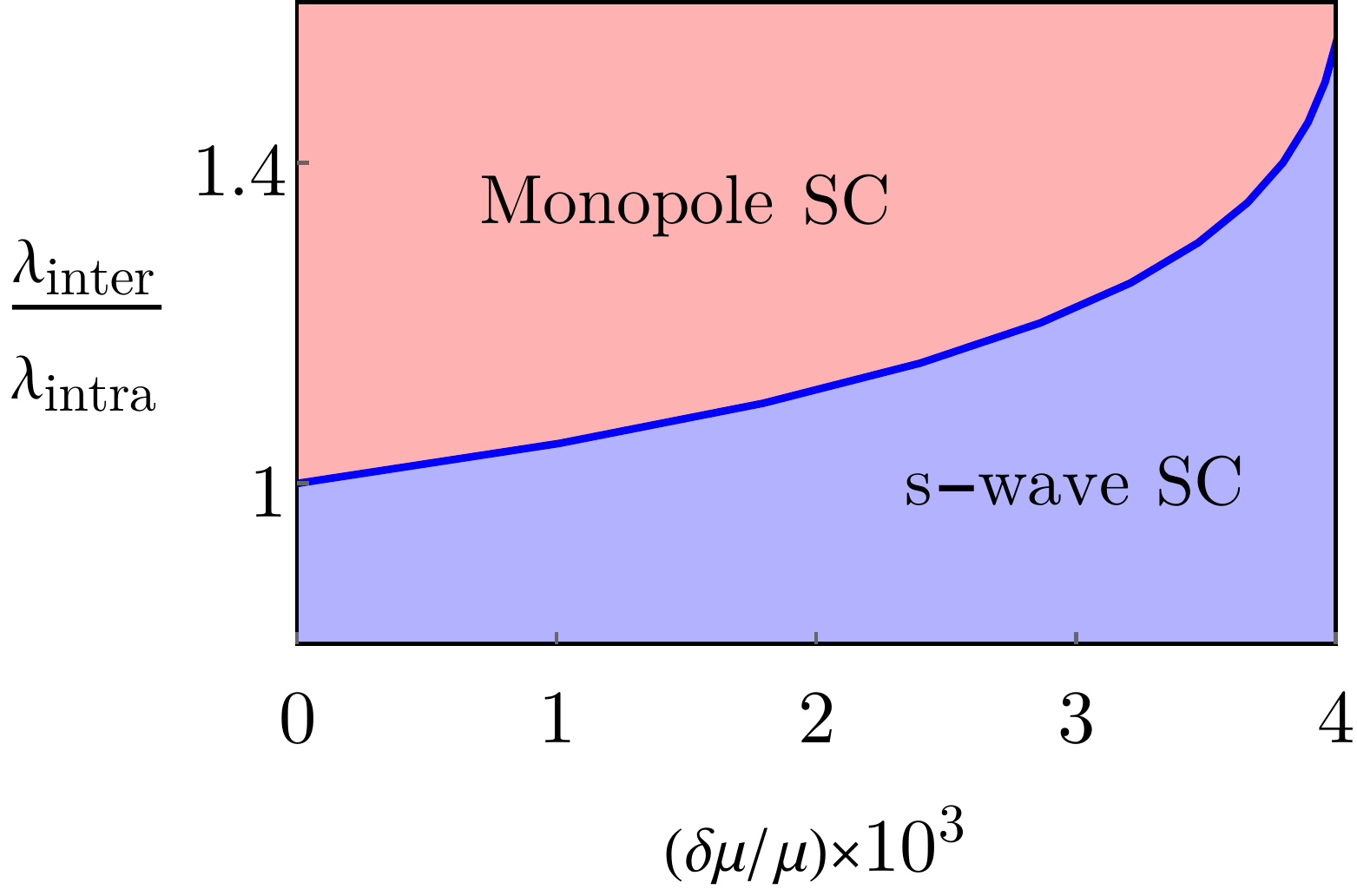}
    \caption{Phase diagram of a Weyl superconductor with the difference between the chemical potentials $\delta\mu$ between the nodes (for $\lambda_{intra}=0.2$). The s-wave is more favorable in this case. The phase diagram is obtained from the behavior of the phase boundary close to the critical temperature from Eqs.~\eqref{eq:deltamu1} and~\eqref{eq:deltamu2}. }
    \label{fig:deltamu}
\end{figure}

To include the effect of the inversion symmetry breaking, we take different chemical potentials at the two nodes and consider the system close to the critical temperature $T_c$. In the symmetric case, $\mu=0$, the finite-temperature phase boundary is given by the condition $\lambda_{intra}=\lambda_{inter}$. When the inversion symmetry is broken, the value of the effective inter-FS pairing potential at the phase boundary, however, increases, implying that the intra-FS $s$-wave superconductor becomes more favorable, as shown in Fig.~\ref{fig:deltamu}.

\subsection{Monopole versus $p-$wave pairing state}
\label{subsection:monopolevspwave}

Let us now consider the $p$-wave case $p_\pm=p_x\pm i p_y$, and $p_z$, defined by $\Delta_{p_z} = \Delta_{p_z,0}\cos\theta$, $\Delta_{p_\pm} = \Delta_{p_\pm,0}e^{\pm i\phi}\sin\theta $. For $p_z$,
we obtain the same behavior as for the $s$-wave case because this harmonic is independent of the azimuthal angle $\phi$ while the monopole harmonic is $\phi$-dependent, and the corresponding $\theta-$dependent form factors are different.
On the other hand, using Eq.~\eqref{eq:BCS-equal}, for both $p_\pm$ pairings we arrive at the BCS equation
\begin{align}
\lambda_{p_\pm}^{-1} &= 2\pi\int_{0}^{\pi}\frac{d\theta}{4\pi}\sin^3\theta
\left[
2\ln(2\omega_D) - 2\ln(\sin\theta) \right.\nonumber\\
&-\left. 2\ln|\Delta_{p_\pm,0}^2 - \bar{\Delta}_0^2|
\right]\nonumber\\
&=
\frac{8}{3}\ln(2\omega_D) - \frac{5}{9}\left(
5 - \ln 64
\right) - \ln|\Delta_{p_\pm,0}^2 - \bar{\Delta}_0^2|
\nonumber\\
\lambda_{intra}^{-1} &=
\frac{8}{3}\ln(2\omega_D) - \frac{5}{9}\left(
5 - \ln 64
\right) - \ln|\Delta_{p_\pm,0}^2 - \bar{\Delta}_0^2|
\end{align}

In this case, coexistence will arise only for $\lambda_{p_\pm} = \lambda_{intra}$, as shown in Fig.~\ref{fig:phasediagram}(b). Otherwise, the dominant phase
will correspond to a larger coupling. After defining the effective coupling by $\lambda_{eff}\equiv \lambda_{p_\pm} = \lambda_{intra}$ and solving the above self-consistent gap equations, we obtain
\begin{eqnarray}
|\Delta_{p_\pm,0}^2 - \bar{\Delta}_0^2| = 2 \omega_D e^{-\frac{1}{12}(5 - \ln 64)} e^{-\frac{3}{8}\lambda_{eff}^{-1}}.
\end{eqnarray}
Ultimately, even in this fine tuned situation, when fluctuation effects are accounted for, we expect that the phases repel each another due to their incompatible topological structure: while the monopole SC features a double vortex coming from individual Fermi surfaces FS$_\pm$, the $p$-wave harmonic picks up a vortex-antivortex pair at each of them.

\begin{figure*}[t!]
    \centering
    \includegraphics[width=0.6\textwidth]{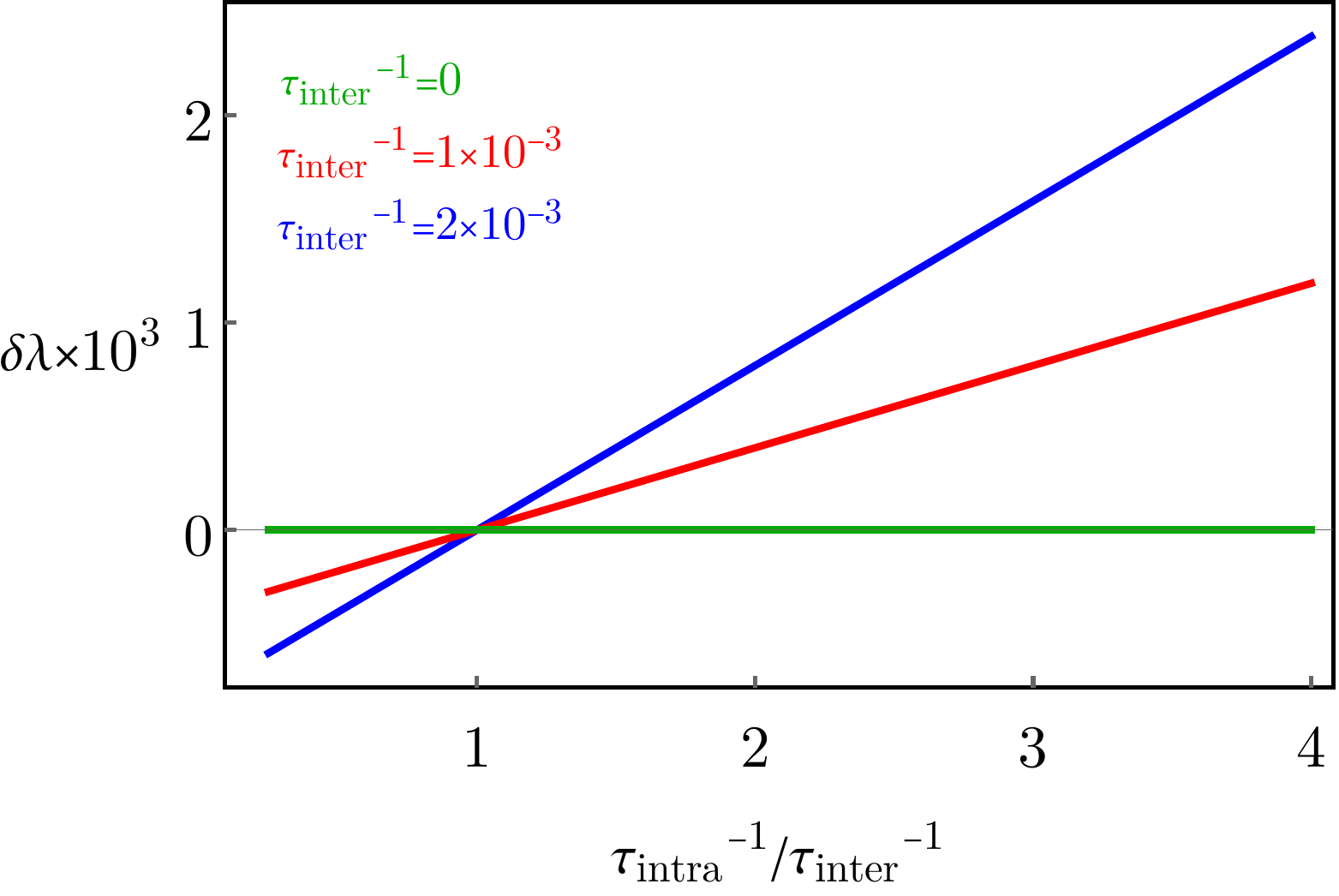}
    \caption{The ratio of the inter- and intra-Fermi surface couplings at the phase boundary (two critical temperatures equal) between the s-wave and the monopole superconductor in the presence of inter- and intra-Fermi surface scattering by the point impurities. The respective scattering times are $\tau_{inter}$ and $\tau_{intra}$ (in units of $1/\mu$). The change of the phase boundary shows that the s-wave pairing is more stable. Here, $\delta\lambda=\lambda_{inter}/\lambda_{intra}-1$, and $\lambda_{inter}=0.13$. }
    \label{fig:taus}
\end{figure*}

\section{Impurity scattering}.
\label{sec:scattering}

Let us now consider the effect of scattering by randomly distributed, non-magnetic impurities, with a concentration $n_{imp}$, on the superconducting instabilities. Within the first Born approximation, this can be captured through an averaged self-energy matrix of the form \cite{Rammer-86}
\begin{eqnarray}
\hat{\Sigma}(\omega_n,\mathbf{q}) = n_{imp}\sum_{\mathbf{q}_1}
\hat{W}_{\mathbf{q},\mathbf{q}_1}\hat{\mathcal{G}}(\omega_n,\mathbf{q}_1)\hat{W}_{\mathbf{q}_1,\mathbf{q}},
\label{eq_selfenergy}
\end{eqnarray}
where $\hat{G}(\omega_n,\mathbf{q})$ is the fully dressed Green's function matrix, arising from the solution of the Dyson equation
\begin{eqnarray}
\hat{\mathcal{G}}^{-1}(\omega_n,\mathbf{q}) = \hat{\mathcal{G}}^{-1}_0(\omega_n,\mathbf{q}) - \hat{\Sigma}(\omega_n,\mathbf{q}).
\label{eq_Dyson}
\end{eqnarray}

We assume, for simplicity, that  the scattering potential matrix is momentum independent, and reads as
\begin{eqnarray}
\hat{W}_{\mathbf{q},\mathbf{q}_1} = u\hat{\tau}_1\otimes\hat{\eta}_0 + v\hat{\tau}_0\otimes\hat{\eta}_3,
\label{eq_W}
\end{eqnarray}
where $u$ and $v$ are, respectively, the inter- and intra-FS scattering strengths, and we assume a Gaussian disorder distribution.
We solve the Dyson equation by taking the ansatz for the fully dressed Green's function so that it retains the same structure as that
without the disorder, but with renormalized parameters $\omega_{n,R}$, $\bar{\Delta}(\mathbf{q})_R$, $\bar{\Delta}_{0,R}$ and $\bar{\xi}_{n,q,R}$.
In the following, we neglect the asymmetry between the chemical potentials at the two Weyl points, and thus set $\delta\mu = 0$.  To the leading order in the impurity strength,
we obtain the renormalized parameters in the form (see Appendix~\ref{app:impurityscattering} for details)
\begin{eqnarray}
{\omega}_{n,R} &=& \omega_n \left( 1 + \frac{1}{2}\left(\tau_{intra}^{-1} + \tau_{inter}^{-1} \right)|\omega_n|^{-1} \right)\nonumber\\
{\Delta}_{0,R} &=& \Delta_0\left( 1 + \frac{1}{2}\left(\tau_{intra}^{-1} - \tau_{inter}^{-1} \right)|\omega_n|^{-1} \right),
\end{eqnarray}
while the band dispersion and the monopole pairing gap function remain unrenormalized. Here, we defined
the inverse inter-FS scattering time $\tau_{inter}^{-1} = 2\pi n_{imp} u^2 \rho(\mu)$, with analogous notation for the intra-FS scattering.

The gap equation for the s-wave  pairing in the weak coupling limit, $\omega_D \gg T_c$  then reads
\begin{equation}
\frac{T_c}{\lambda_{intra}} =  \ln\left(\frac{\Gamma_c}{2\pi T_c} \right)
- \psi\left(\frac{1}{2} + \frac{\tau_{inter}^{-1}}{2\pi T_c}  \right),
\end{equation}
where $\Gamma_c$ is an upper cutoff for the Matsubara frequency sum and $\psi(x)$ is the digamma function. When only the intra-FS scattering is present, the critical temperature remains unchanged, consistent with Anderson's theorem~\cite{Anderson-theorem}, and its generalized version for unconventional pairing states in terms of the superconducting fitness~\cite{Ramires-PRB2018,Andersen-SciAdv2020}. Notice that the density-wave nature of this superconducting order does not play a role, since the  disorder preserves translational symmetry on average.

For the monopole superconductor, on the other hand, because of the form of its projection on the FS, the gap does not renormalize.  Consequently, the effect of both intra- and inter-FS disorders is to lower its critical temperature,
\begin{eqnarray}\label{eq:disorder-monopole}
\frac{T_c}{\lambda_{inter}} = \ln\left(\frac{\Gamma_c}{2\pi T_c} \right)
- \psi\left( \frac{1}{2} + \frac{\tau_{intra}^{-1} + \tau_{inter}^{-1}}{4\pi T_c} \right).
\end{eqnarray}
Notice that  both types of disorder \emph{anticommute} with the pairing matrix for the monopole SC, as can be directly checked from Eq.~\eqref{eq_H1}. Therefore, the superconducting fitness function is non-vanishing for either of them, consistent with the correction to $T_c$ given by Eq.~\eqref{eq:disorder-monopole}.

To illustrate the competition of the two superconducting phases when  the intra-FS scattering is turned on, we plot the ratio between the inter- and intra-FS pairing interactions, $\delta\lambda=\lambda_{inter}/\lambda_{intra}-1$ at
the phase boundary (the two critical temperatures are equal) as a function of the intra-FS inverse scattering time, shown in Fig.~\ref{fig:taus}. The intra-FS scattering suppresses the monopole SC, since as this scattering increases, the phase boundary moves toward larger values of the inter-FS pairing strength. A similar behavior is observed when the inter-FS scattering is tuned for a fixed intra-FS disorder.

\section{Discussion and Outlook}
\label{sec:discussion}

To summarize, we here demonstrated that the monopole and a conventional spherical harmonic SCs quite generically coexist, while the repulsion can take place when the absolute value of the monopole charge matches the angular momentum quantum number of the spherical harmonic. We illustrated this general principle on the particular examples of finite-momentum $s-$ and $p-$wave pairings competing with the monopole SC, which, respectively,  coexist and repel.  We showed that the $s$-wave pairing is more stable both for unequal chemical potentials at the nodes,  and  in  the  presence  of  point-like charged  impurities.
Close to the phase boundary, the system  features gapless modes at the interface of the topologically nontrivial monopole harmonic and the trivial $s$-wave superconducting domains, providing an experimental signature of the monopole SC.

In spite of many realized Weyl metals, the signatures of the Weyl superconductivity were only recently reported  in UTe$_2$~\cite{Hayes-arxiv2020}. Particularly relevant in this context is the observation that the superconducting state is a time-reversal symmetry breaking two-component spin-triplet order parameter, which as such may feature a monopole component, but the nature of the order parameter is still an open question.

Our work should motivate further studies of the monopole harmonic SCs, such as  their competition with the insulating instabilities, particularly with those displaying the monopole structure~\cite{Bobrow-PRR2020}. Finally, observable consequences of these exotic states beyond the surface Majorana modes are yet to be explored,  for instance, impurity resonances~\cite{balatsky-RMP2006}.


\section{Acknowledgment}
We thank Bitan Roy for insightful comments. This work was supported by Fondecyt Grants 1190361 and 1200399, and by ANID PIA/Anillo ACT192023. V.J. acknowledges the
support of the Swedish Research Council (VR 2019-04735).

\appendix

 \section{Details of the model}
 \label{app:Model}
 We start by considering the model introduced in Refs.~\cite{Li-Hadane-PRL2018,Sun-arxiv2019}, which describes the mean-field Cooper pairing between quasiparticle excitations at two Fermi surfaces (FSs) enclosing the two nodal points
 in a Weyl semimetal. For this model, the effective Hamiltonian
 is
 \begin{eqnarray}
\hat{H} = \hat{H}_\text{Weyl} + \hat{H}_{\Delta}.
 \label{eq_1}
 \end{eqnarray}
 Here,  $\hat{H}_\text{Weyl} = \sum_{\mathbf{k}}\hat{c}_{\mathbf{k}}^{\dagger}h(\mathbf{k})\hat{c}_{\mathbf{k}}$, where the matrix

 \begin{align}
h(\mathbf{k}) &= t \sin k_x \hat{\sigma}_x+ t \sin k_y\hat{\sigma}_y\nonumber\\
&+  t (2 - \cos k_x- \cos k_y - \cos k_z
+ \cos K_0 )\hat{\sigma}_z - \mu,
 \label{eq_2}
 \end{align}
 and $\mu>0$ is the chemical potential.

 The band structure in Eq.~(\ref{eq_2}) possesses two Weyl nodes along the $k_z$-direction, given by $\zeta\mathbf{K}_0 = \left(0,0,\zeta K_0 \right)$,
 each with opposite topological charge $\mathcal{C}_{\zeta} = \zeta = \pm 1$. We consider only isotropic nodes by choosing $t = v_F$, and $K_0 = \pi/2$. Therefore, we write the effective Hamiltonian describing Weyl quasiparticles in the vicinity of each node $\zeta\mathbf{K}_0$ as $\hat{H}_{Weyl} = \sum_{\zeta = \pm,\mathbf{q}}\hat{c}^{\dagger}_{\zeta \mathbf{K}_0 + \mathbf{q}}\hat{h}_{\zeta}(\mathbf{q})\hat{c}_{\zeta\mathbf{K}_0 + \mathbf{q}}$, with
 \begin{eqnarray}
\hat{h}_{\zeta}(\mathbf{q}) = v_F\left( \hat{\sigma}_x q_x + \hat{\sigma}_y q_y + \zeta \hat{\sigma}_z q_z  \right) - \mu,
 \end{eqnarray}
which we use in the main text.

\section{~Bogoliubov-de Gennes finite temperature field-theory}
\label{app:fieldtheory}
From the effective mean-field Bogoliubov-de Gennes Hamiltonian in Eq.~(\ref{eq_H1}), we can construct a finite temperature field theory,
in terms of the Grassmann four-component fermion fields
in the Nambu basis
$\Psi_{\mathbf{q}}$ and $\Psi^{\dagger}_{\mathbf{q}}$. The corresponding partition function is given by the functional integral
\begin{eqnarray}
Z_0 = \int\mathcal{D}\Psi_{\mathbf{q}}^{\dagger}\mathcal{D}\Psi_{\mathbf{q}}\, e^{-S\left[ \Psi_{\mathbf{q}}^{\dagger},\Psi_{\mathbf{q}} \right]},
\end{eqnarray}
with the action in compactified Euclidean time $0 \le \tau \le \beta$ (for $\beta = 1/T$)
\begin{equation}
S\left[ \Psi_{\mathbf{q}}^{\dagger},\Psi_{\mathbf{q}} \right]
= \int_{0}^{\beta}d\tau\,\Psi_{\mathbf{q}}^{\dagger}\left[\hat{\tau}_0\otimes\hat{\eta}_0\frac{\partial}{\partial\tau} + \hat{H}_{BdG}(\mathbf{q})  \right]\Psi_{\mathbf{q}}.
\end{equation}
Therefore, the corresponding matrix Green's function satisfies the differential equation
\begin{eqnarray}
\left[\frac{\partial}{\partial\tau} + \hat{H}_{BdG}(\mathbf{q})  \right]\hat{\mathcal{G}}_0(\tau,\mathbf{q}) = \delta(\tau),
\label{eq:Green_1}
\end{eqnarray}
which in the Matsubara frequency space assumes the form
\begin{eqnarray}
\hat{\mathcal{G}}_0(\omega_n,\mathbf{q}) = \int_0^{\beta}d\tau e^{i\omega_n\tau}\,\hat{\mathcal{G}}_0(\tau,\mathbf{q}),
\end{eqnarray}
with $\omega_n = (2n + 1)\pi/\beta$ for $n\in \mathbb{Z}$.
We now solve Eq.~(\ref{eq:Green_1}) to obtain
\begin{eqnarray}
\hat{\mathcal{G}}_0(\omega_n,\mathbf{q}) &=& \left[-i\omega_n\hat{\tau}_0\otimes\hat{\eta}_0 + \hat{H}_{BdG}(\mathbf{q}) \right]^{-1}.
\label{eq:Green_2}
\end{eqnarray}
\section{Clean limit}
\label{app:cleanlimit}
We first analyze the system in the absence of impurities to find the critical temperature in this clean limit.
We first rewrite the Hamiltonian in Eq.~\eqref{eq:HBdG2} in terms of the 16 $\Gamma-$matrices $\Gamma_{ij} = \hat{\tau}_i\otimes\hat{\eta}_j$, $i,j=0,1,2,3$,
\begin{equation}
\hat{H}_{BdG}(\mathbf{q}) = \sum_{\alpha=1}^3 a_{0\alpha}\Gamma_{0\alpha}+a_{11}\Gamma_{11}+a_{12}\Gamma_{12}+a_{33}\Gamma_{33}
\label{eq:HGam}
\end{equation}
where we defined the coefficients
\begin{eqnarray}
a_{03} &=& \bar{\xi}_{q},\,\,\, a_{01} = \Re \Delta_0,\,\,\,a_{02} = - \Im\Delta_0\nonumber\\
a_{11} &=& \Re\tilde{\Delta}(\mathbf{q}),\,\,\,a_{12} = -\Im\tilde{\Delta}(\mathbf{q}),
\,\,\,a_{33} = \frac{\delta\mu}{2}.
\end{eqnarray}
The Green's function is obtained by calculating
the inverse matrix in Eq.~(\ref{eq:Green_2}), as follows
\begin{equation}
\hat{\mathcal{G}}_0(\omega_n,\mathbf{q}) = \left[i\omega_n\hat{\tau}_0\otimes\hat{\eta}_0 + \hat{H}_{BdG}(\mathbf{q})\right]\left[\omega_n^2 + \hat{H}_{BdG}(\mathbf{q})^2\right]^{-1}.
\end{equation}
From Eq.~(\ref{eq:HGam}) and using the anticommutation relations of the Pauli matrices, we obtain
\begin{widetext}
\begin{equation}
\hat{H}_{BdG}(\mathbf{q})^2  = \Gamma_{00}b^2
+ 2 \left(a_{01} a_{11} \Gamma_{01} \Gamma_{11} + a_{02} a_{12} \Gamma_{02} \Gamma_{12} + a_{03} a_{33} \Gamma_{03} \Gamma_{33}   \right),
\end{equation}
\end{widetext}
where
\begin{equation}
b^2=\sum_{\alpha=1}^3a_{0\alpha}^2+a_{11}^2+a_{12}^2+a_{33}^2.  \end{equation}
After some straightforward algebra, we obtain the final result:
\begin{equation}\label{eq:G0full}
\hat{\mathcal{G}}_0(\omega_n,\mathbf{q}) = i\omega_n \frac{\bar{E}_q^2 \Gamma_{00} - 2 B_q \Gamma_{10} + \delta\mu\bar{\xi}_q \Gamma_{30}}{\bar{E}_q^4 - 4 B_q^2 - \delta\mu^2 \bar{\xi}_q^2} + \sum_{j=1}^{12}\hat{A}_j.
\end{equation}
Here, we define the parameters
\begin{eqnarray}
\bar{E}_q^2 &=& \bar{\xi}_q^{2} + |\Delta_0|^2 + |\tilde{\Delta}(\mathbf{q})|^2 + \frac{\delta\mu^2}{4} + \omega_n^2\nonumber\\
B_q &=& \Re\Delta_0 \Re\tilde{\Delta}(\mathbf{q}) + \Im\Delta_0 \Im\tilde{\Delta}(\mathbf{q}),
\end{eqnarray}
and the set of matrices
\begin{align}
\hat{A}_1 &= - \frac{\bar{\xi}_q \bar{E}_q^2 + \bar{\xi}_q \delta\mu^2}{\bar{E}_q^4 - 4 B_q^2 - \delta\mu^2 \bar{\xi}_q^2}\hat{\tau}_0\otimes\hat{\eta}_3 \equiv \tilde{a}_{03} \Gamma_{03}\nonumber\\
\hat{A}_2 &=  \frac{2 B_q \Re\tilde{\Delta}(\mathbf{q}) - \bar{E}_q^2 \Re\Delta_0}{\bar{E}_q^4 - 4 B_q^2 - \delta\mu^2 \bar{\xi}_q^2}\hat{\tau}_0\otimes\hat{\eta}_1 \equiv \tilde{a}_{01} \Gamma_{01}\nonumber\\
\hat{A}_3 &=  \frac{\bar{E}_q^2\Im\Delta_0 - 2 B_q \Im \tilde{\Delta}(\mathbf{q})}{\bar{E}_q^4 - 4 B_q^2 - \delta\mu^2 \bar{\xi}_q^2}\hat{\tau}_0\otimes\hat{\eta}_2 \equiv \tilde{a}_{02} \Gamma_{02}\nonumber\\
\hat{A}_4 &=  \frac{2 B_q \Re\Delta_0 - \bar{E}_q^2 \Re\tilde{\Delta}(\mathbf{q})}{\bar{E}_q^4 - 4 B_q^2 - \delta\mu^2 \bar{\xi}_q^2}\hat{\tau}_1\otimes\hat{\eta}_1 \equiv \tilde{a}_{11} \Gamma_{11}\nonumber\\
\hat{A}_5 &=  \frac{\bar{E}_q^2\Im\tilde{\Delta}(\mathbf{q}) - 2 B_q \Im\Delta_0}{\bar{E}_q^4 - 4 B_q^2 - \delta\mu^2 \bar{\xi}_q^2}\hat{\tau}_1\otimes\hat{\eta}_2 \equiv \tilde{a}_{12} \Gamma_{12}\nonumber\\
\hat{A}_6 &=  \frac{2 B_q \bar{\xi}_q}{\bar{E}_q^4 - 4 B_q^2 - \delta\mu^2 \bar{\xi}_q^2}\hat{\tau}_1\otimes\hat{\eta}_3 \equiv \tilde{a}_{13} \Gamma_{13}\nonumber
\end{align}
\begin{align}
\hat{A}_7 &=  -\frac{\delta\mu\left( \bar{\xi}_q^2 + \bar{E}_q^2\right)}{\bar{E}_q^4 - 4 B_q^2 - \delta\mu^2 \bar{\xi}_q^2}\hat{\tau}_3\otimes\hat{\eta}_3 \equiv \tilde{a}_{33} \Gamma_{33}\nonumber\\
\hat{A}_8 &=  \frac{2 i \delta\mu B_q}{\bar{E}_q^4 - 4 B_q^2 - \delta\mu^2 \bar{\xi}_q^2}\hat{\tau}_2\otimes\hat{\eta}_3 \equiv \tilde{a}_{23} \Gamma_{23}\nonumber\\
\hat{A}_9 &=  -\frac{\delta\mu \bar{\xi}_q\Re\Delta_0}{\bar{E}_q^4 - 4 B_q^2 - \delta\mu^2 \bar{\xi}_q^2}\hat{\tau}_3\otimes\hat{\eta}_1 \equiv \tilde{a}_{31} \Gamma_{31}\nonumber\\
\hat{A}_{10} &=  \frac{\delta\mu \bar{\xi}_q\Im\Delta_0}{\bar{E}_q^4 - 4 B_q^2 - \delta\mu^2 \bar{\xi}_q^2}\hat{\tau}_3\otimes\hat{\eta}_2 \equiv \tilde{a}_{32} \Gamma_{32}\nonumber\\
\hat{A}_{11} &=  \frac{i \delta\mu \bar{\xi}_q\Re\tilde{\Delta}(\mathbf{q})}{\bar{E}_q^4 - 4 B_q^2 - \delta\mu^2 \bar{\xi}_q^2}\hat{\tau}_2\otimes\hat{\eta}_1 \equiv \tilde{a}_{21} \Gamma_{21}\nonumber\\
\hat{A}_{12} &=  -\frac{i \delta\mu \bar{\xi}_q\Im\tilde{\Delta}(\mathbf{q})}{\bar{E}_q^4 - 4 B_q^2 - \delta\mu^2 \bar{\xi}_q^2}\hat{\tau}_2\otimes\hat{\eta}_2 \equiv \tilde{a}_{22} \Gamma_{22}.
\end{align}
Taking these definitions into account, we write the Green's function in the block-matrix form as
\begin{eqnarray}
\hat{\mathcal{G}}_0(\omega_n,\mathbf{q}) = \left[\begin{array}{cc}\hat{G}_{0}^{--} & \hat{G}_0^{-+}\\\hat{G}_0^{+-}&\hat{G}_0^{++}\end{array}\right].
\label{eq:Gomat}
\end{eqnarray}
In particular, we are interested in the correlators
\begin{eqnarray}
\langle \hat{\alpha}_{-}(\mathbf{q})\hat{\alpha}_{+}(-\mathbf{q})\rangle &=& \left[\hat{G}_0^{-+} \right]_{21} = \tilde{a}_{11} - i\tilde{a}_{21} - i\left( \tilde{a}_{12} - i \tilde{a}_{22} \right)\nonumber\\
&=& \frac{2 B_q \Delta_0 - \bar{E}_q^2 \tilde{\Delta}(\mathbf{q}) + \delta\mu \bar{\xi}_q \tilde{\Delta}(\mathbf{q})}{\bar{E}_q^4 - 4 B_q^2 - \delta\mu^2 \bar{\xi}_q^2},
\end{eqnarray}
and
\begin{eqnarray}
\langle \hat{\alpha}_{-}(\mathbf{q})\hat{\alpha}_{-}(-\mathbf{q})\rangle &=& \left[\hat{G}_0^{--} \right]_{21} = \tilde{a}_{01} + \tilde{a}_{31} - i\left( \tilde{a}_{02} + \tilde{a}_{32} \right)\nonumber\\
&=& \frac{2 B_q \tilde{\Delta}(\mathbf{q}) - \bar{E}_q^2 \Delta_0 - \delta\mu \bar{\xi}_q \Delta_0}{\bar{E}_q^4 - 4 B_q^2 - \delta\mu^2 \bar{\xi}_q^2}.
\end{eqnarray}
\subsection{BCS gap equations}
Let us now consider the self-consistent BCS gap equations arising from the clean system previously discussed for the linear order in pairing amplitudes. Since we assume inter-FS and intra-FS pairings, we then obtain a pair of coupled self-consistent BCS gap equations
\begin{align}
&\Delta_0 = -T\sum_{\mathbf{q}',\omega_n}V_{intra}(\mathbf{q},\mathbf{q}')
\langle\hat{\alpha}_{-}(\mathbf{q}')\hat{\alpha}_{-}(-\mathbf{q}')\rangle\nonumber\\
&= \sum_{\mathbf{q}'}V_{intra}(\mathbf{q}')\left\{
\left(\frac{\Delta_0}{2} - \frac{2 B_{q'} \tilde{\Delta}(\mathbf{q}') - \delta\mu \bar{\xi}_{q'}\Delta_0}{\sqrt{4 B_{q'}^2 + \delta\mu^2\bar{\xi}_{q'}^2}}\right)
{\mathcal T}_\beta(\gamma_{\mathbf{q}'})\right.
\nonumber\\
&\left.+
\left(\frac{\Delta_0}{2} + \frac{2 B_{q'} \tilde{\Delta}(\mathbf{q'}) - \delta\mu \bar{\xi}_{q'}\Delta_0}{\sqrt{4 B_{q'}^2 + \delta\mu^2\bar{\xi}_{q'}^2}}\right)
{\mathcal T}_\beta(\Gamma_{\mathbf{q}'})\right\},
\label{eq:BCS_intra}
\end{align}

\begin{align}
&\tilde{\Delta}(\mathbf{q}) = - T \sum_{\mathbf{q}',\omega_n}V_{inter}(\mathbf{q},
\mathbf{q}')\langle \hat{\alpha}_{-}(\mathbf{q}')\hat{\alpha}_{+}(-\mathbf{q}') \rangle\nonumber\\
&= \sum_{\mathbf{q}'}V_{inter}(\mathbf{q},\mathbf{q}')\left\{
\left(\frac{\tilde{\Delta}(\mathbf{q}')}{2} - \frac{2 B_{q'} \Delta_0 + \delta\mu \bar{\xi}_{q'}\tilde{\Delta}(\mathbf{q}')}{\sqrt{4 B_{q'}^2 + \delta\mu^2\bar{\xi}_{q'}^2}}\right)
{\mathcal T}_\beta(\gamma_{\mathbf{q}'})\right.\nonumber\\
&+\left.
\left(\frac{\tilde{\Delta}(\mathbf{q}')}{2} + \frac{2 B_{q'} \Delta_0 + \delta\mu \bar{\xi}_{q'}\tilde{\Delta}(\mathbf{q}')}{\sqrt{4 B_{q'}^2 + \delta\mu^2\bar{\xi}_{q'}^2}}\right)
{\mathcal T}_\beta(\Gamma_{\mathbf{q}'})\right\}.
\label{eq:BCS_inter}
\end{align}
Here, ${\mathcal T}_\beta(x)\equiv\tanh(\beta x/2)/x$.
To arrive at these expressions, we performed a partial fraction decomposition by factoring the denominator in the correlators as follows
\begin{widetext}
\begin{equation}
\bar{E}_q^4 - 4 B_q^2 - \delta\mu^2\bar{\xi}_q^2 =
\left(\bar{E}_q^2 + \sqrt{4 B_q^2 + \delta\mu^2\bar{\xi}_q^2} \right)\left(\bar{E}_q^2 - \sqrt{4 B_q^2 + \delta\mu^2\bar{\xi}_q^2} \right)\nonumber= \left(\omega_n^2 + \Gamma_q^2 \right)\left(\omega_n^2 +\gamma_q^2
\right),
\end{equation}
\end{widetext}
where we defined the parameters
\begin{align}
&\Gamma_q^2 = \bar{\xi}_q^2 + |\Delta_0|^2 + |\tilde{\Delta}(\mathbf{q})|^2 + \frac{\delta\mu^2}{4}
+ \sqrt{4 B_q^2 + \delta\mu^2\bar{\xi}_q^2},\nonumber\\
&\gamma_q^2 = \bar{\xi}_q^2 + |\Delta_0|^2 + |\tilde{\Delta}(\mathbf{q})|^2 + \frac{\delta\mu^2}{4}
- \sqrt{4 B_q^2 + \delta\mu^2\bar{\xi}_q^2}.
\end{align}
The Matsubara sums were performed by partial fraction decomposition, followed by the application of the basic identity (for $c\in \mathbb{R}$)
\begin{equation}
T\sum_{\omega_n}\frac{1}{\omega_n^2 + c^2} = \frac{\tanh\left(\frac{c}{2T}\right)}{c}.
\end{equation}
Let us now calculate the critical temperature $T_c = \beta_c^{-1}$ by imposing the condition $\Delta_0\rightarrow0$, $\tilde{\Delta}(\mathbf{q})\rightarrow 0$ in Eqs.~\eqref{eq:BCS_intra} and~\eqref{eq:BCS_inter}. In this limit, we have $B_q \rightarrow 0$, and
hence the parameters defined above reduce to
\begin{align}
&\Gamma_q \rightarrow \bar{\xi}_q + \frac{\delta\mu}{2} = v_F q - \mu^{-},\nonumber\\
&\gamma_q \rightarrow \bar{\xi}_q - \frac{\delta\mu}{2} = v_F q - \mu^{+}.
\end{align}
The forms of the spherical harmonic and the monopole SC 
order parameters for the intra- and inter-FS pairings dictate  the angular dependence of the pairing potentials for each channel
\begin{eqnarray}
V_{intra}(\mathbf{q},\mathbf{q}') &=& V_0 Y_{l,m}(\theta_q,\phi_q)Y_{l,m}^{*}(\theta_{q'},\phi_{q'})\nonumber\\
V_{inter}(\mathbf{q},\mathbf{q}') &=& \tilde{V}_0 Y_{-1,1,0}(\theta_q,\phi_{q})Y_{-1,1,0}^{*}(\theta_{q'},\phi_{q'}).\nonumber\\
\label{eq:potential}
\end{eqnarray}
Let us first consider the simplest case of an $s$-wave intra-nodal pairing, represented in Eq.~(\ref{eq:potential}) by the spherical harmonic $l = m = 0$, which is just a constant. Under this assumption, the coupled BCS equations, at the critical temperature $T_c$, reduce to the simpler expressions
\begin{align}
 V_0^{-1} &= \int \frac{d^3 q}{(2\pi)^3}\left[
\frac{(1 + {\rm{sgn}}(\delta\mu\bar{\xi}_q))}{2}
\frac{\tanh(\beta_c(\bar{\xi}_q - \delta\mu/2))}{\bar{\xi}_q - \delta\mu/2}\right.\nonumber\\
&\left.+ \frac{(1 - {\rm{sgn}}(\delta\mu\bar{\xi}_q))}{2}
\frac{\tanh(\beta_c(\bar{\xi}_q + \delta\mu/2))}{\bar{\xi}_q + \delta\mu/2}
\right]
\label{eq:BCS2}
\end{align}
\begin{eqnarray}
\kappa^{-1} &=& \int \frac{d^3 q}{(2\pi)^3}\left[
\frac{(1 - {\rm{sgn}}(\delta\mu\bar{\xi}_q))}{2}
\frac{\tanh(\beta_c(\bar{\xi}_q - \delta\mu/2))}{\bar{\xi}_q - \delta\mu/2}\right.\nonumber\\
&&\left.+ \frac{(1 + {\rm{sgn}}(\delta\mu\bar{\xi}_q))}{2}
\frac{\tanh(\beta_c(\bar{\xi}_q + \delta\mu/2))}{\bar{\xi}_q + \delta\mu/2}
\right].
\label{eq:BCS1}
\end{eqnarray}
Here, we defined the coefficient
\begin{eqnarray}
\kappa = \frac{\tilde{V}_0}{4\pi}\int d\Omega_{q} |Y_{-1,1,0}(\theta_q,\phi_q)|^2.
\label{eq:kappa}
\end{eqnarray}
It is convenient to change integration variables in the momentum integrals defined above, by introducing the density of states
\begin{widetext}
\begin{equation}
\rho(\xi) =\int \frac{d^3q}{(2\pi)^3} \delta(\xi + \mu -  v_F q) = \int \frac{d\Omega_q}{(2\pi)^3}\int_{0}^{\infty}
dq q^2 \delta(\xi + \mu - v_F q)=
\frac{4\pi}{(2\pi)^3}\frac{(\xi + \mu)^2}{( v_F)^3},
\label{eq:DOS}
\end{equation}
to express them as energy integrals, within
a symmetric interval $-\omega_D \le \xi \le  \omega_D$, centered at the chemical
potential and bounded by the physical cutoff provided by the phonon Debye frequency $\omega_D$. With these considerations for $\mathcal{F}(\Omega_q,\xi_q)$, which is  an arbitrary function of $\mathbf{q}$, and $d\Omega_q = d\theta d\phi\sin\theta\cos\phi$
the differential solid angle, we adopt the prescription
\begin{equation}
\int\frac{d^3q}{(2\pi)^3}\mathcal{F}(\Omega_q,\xi_q) = \int \frac{d\Omega_q}{4\pi} \int_{-\omega_D}^{\omega_D}d\xi \rho(\xi)\mathcal{F}(\Omega_q,\xi)\sim \rho(\mu)\int \frac{d\Omega_q}{4\pi}\int_{-\omega_D}^{\omega_D}
d\xi\mathcal{F}(\Omega_q,\xi).
\label{eq:dos-integration}
\end{equation}
\end{widetext}
By converting the momentum integral into the one over energy through the density of states, as described above,  the Eqs.~(\ref{eq:BCS2}) and~(\ref{eq:BCS1}) reduce to
\begin{eqnarray}
\lambda_{intra}^{-1} &=& \int_{-\omega_D}^{\omega_D} d\xi \left[
\frac{(1 + {\rm{sgn}}(\delta\mu \xi))}{2}
\frac{\tanh(\beta_c(\xi - \delta\mu/2))}{\xi - \delta\mu/2}\right.\nonumber\\
&&\left.+ \frac{(1 - {\rm{sgn}}(\delta\mu\xi))}{2}
\frac{\tanh(\beta_c(\xi + \delta\mu/2))}{\xi + \delta\mu/2}
\right],
\label{eq:deltamu1}
\end{eqnarray}
\begin{eqnarray}
\lambda^{-1}_{inter} &=& \int_{-\omega_D}^{\omega_D} d\xi\left[
\frac{(1 - {\rm{sgn}}(\delta\mu\xi))}{2}
\frac{\tanh(\beta_c(\xi - \delta\mu/2))}{\xi - \delta\mu/2}\right.\nonumber\\
&&\left.+ \frac{(1 + {\rm{sgn}}(\delta\mu \xi))}{2}
\frac{\tanh(\beta_c(\xi + \delta\mu/2))}{\xi + \delta\mu/2}
\right].
\label{eq:deltamu2}
\end{eqnarray}
Here, we defined the effective couplings
\begin{align}
\lambda_{inter} &= V_0\rho(\mu),\nonumber\\
\lambda_{intra} &= \kappa\rho(\mu),
\label{eq:coupling}
\end{align}
with  the density of states $\rho(\mu)$, as defined in Eq.(\ref{eq:DOS}), evaluated at the average chemical potential.

Let us now consider the $T=0$ BCS gap equation for a more general intra-FS pairing as defined in Eq.~(\ref{eq:potential}). For convenience, we introduce the notation
\begin{align}
\Delta_{inter}(\mathbf{q}) &= \bar{\Delta}_0\, d_{inter}(\theta)\, e^{i(m+q)\phi}\nonumber\\
\Delta_{intra}(\mathbf{q}) &= \Delta_{lm,0}\, d_{intra}(\theta)\, e^{im'\phi},
\end{align}
where for the spherical harmonics we write $Y_{l,m'}(\theta,\phi) = e^{i m'\phi}f_l(\theta)$, and choose $m'>0$ without loss of generality. The form of the monopole harmonic is given by Eq.~\eqref{eq:mono-definition}, which we rewrite as $Y_{q,j,m}=e^{i(m+q)\phi}g(\theta)$, with other indices omitted for clarity, and we set $m+q>0$ since the choice $m+q<0$ is gauge equivalent, as shown below.

The generalization of Eq.(\ref{eq:BCS_intra}) and Eq.(\ref{eq:BCS_inter}), obtained from the Green's function in Eq.~\eqref{eq:G0full}, retaining also nonlinear terms in the pairing amplitudes then reads
\begin{widetext}

\begin{align}
\Delta_{intra}(\mathbf{q}) &= -T\sum_{\mathbf{q}',\omega_n}V_{intra}(\mathbf{q},\mathbf{q}')
\langle\hat{\alpha}_{-}(\mathbf{q}')\hat{\alpha}_{-}(-\mathbf{q}')\rangle\nonumber\\
&= \sum_{\mathbf{q}'}V_{intra}(\mathbf{q},\mathbf{q}')\left\{
\left(\frac{\Delta_{intra}(\mathbf{q}')}{2} - \frac{2 B_{q'} \Delta_{inter}(\mathbf{q}') - \delta\mu \bar{\xi}_{q'}\Delta_{intra}(\mathbf{q}')}{\sqrt{4 B_{q'}^2 + \delta\mu^2\bar{\xi}_{q'}^2}}\right)\frac{\tanh(\beta\gamma_{\mathbf{q'}}/2)}{\gamma_{\mathbf{q'}}}\right.\nonumber\\
&\left.+
\left(\frac{\Delta_{intra}(\mathbf{q}')}{2} + \frac{2 B_{q'} \Delta_{inter}(\mathbf{q}') - \delta\mu \bar{\xi}_{q'}\Delta_{intra}(\mathbf{q}')}{\sqrt{4 B_{q'}^2 + \delta\mu^2\bar{\xi}_{q'}^2}}\right)
\frac{\tanh(\beta\Gamma_{\mathbf{q'}}/2)}{\Gamma_{\mathbf{q'}}}\right\},
\label{eq:BCS_intra2}
\end{align}
\begin{align}
\Delta_{inter}(\mathbf{q}) &= - T \sum_{\mathbf{q}',\omega_n}V_{inter}(\mathbf{q},\mathbf{q}')\langle \hat{\alpha}_{-}(\mathbf{q}')\hat{\alpha}_{+}(-\mathbf{q}') \rangle\nonumber\\
&= \sum_{\mathbf{q}'}V_{inter}(\mathbf{q},\mathbf{q}')\left\{
\left(\frac{\Delta_{inter}(\mathbf{q}')}{2} - \frac{2 B_{q'} \Delta_{intra}(\mathbf{q}') + \delta\mu \bar{\xi}_{q'}\Delta_{inter}(\mathbf{q}')}{\sqrt{4 B_{q'}^2 + \delta\mu^2\bar{\xi}_{q'}^2}}\right)\frac{\tanh(\beta\gamma_{\mathbf{q}'}/2)}{\gamma_{\mathbf{q}'}}\right.\nonumber\\
&\left.+
\left(\frac{\Delta_{inter}(\mathbf{q}')}{2} + \frac{2 B_{q'} \Delta_{intra}(\mathbf{q}') + \delta\mu \bar{\xi}_{q'}\Delta_{inter}(\mathbf{q}')}{\sqrt{4 B_{q'}^2 + \delta\mu^2\bar{\xi}_{q'}^2}}\right)
\frac{\tanh(\beta\Gamma_{\mathbf{q}'}/2)}{\Gamma_{\mathbf{q}'}}\right\}.
\label{eq:BCS_inter2}
\end{align}
with the parameter $B_q$ now written in the form
\begin{equation}
B_q = \Re\Delta_{inter}(\mathbf{q})\Re\Delta_{intra}(\mathbf{q})
+ \Im\Delta_{inter}(\mathbf{q})\Im\Delta_{intra}(\mathbf{q}).
\end{equation}

As in the first example, we are interested in the vicinity of the phase boundary, where
both $|\Delta_{inter}|\ll 1$ and $|\Delta_{intra}|\ll 1$, such that we retain only linear terms in the numerator of Eqs.(\ref{eq:BCS_intra2}) and~(\ref{eq:BCS_inter2}).

After integrating over $|\xi|\le\omega_D$
in Eqs.~(\ref{eq:BCS_intra}) and~(\ref{eq:BCS_inter}), using the identity in Eq.~\eqref{eq:dos-integration}, and the result
\begin{eqnarray}
&&\int_{-\omega_D}^{\omega_D}d\xi\frac{\rho(\xi)}{\sqrt{\xi^2 + |\Delta_{inter}|^2 + |\Delta_{intra}|^2
\pm 2 \left[\Re\Delta_{inter}\Re\Delta_{intra}
+ \Im\Delta_{inter}\Im\Delta_{intra}\right]}}\\
&=& 2\rho(\mu) \left\{
\ln(2\omega_D) - \frac{1}{2}\ln\left[
|\Delta_{inter}|^2 + |\Delta_{intra}|^2
\pm 2 \left(\Re\Delta_{inter}\Re\Delta_{intra}
+ \Im\Delta_{inter}\Im\Delta_{intra}\right)
\right]
\right\},\nonumber
\end{eqnarray}
we obtain the corresponding generalized  zero-temperature BCS gap equations in Eq.~\eqref{eq:BCS-general} in the main text.
\end{widetext}
\section{Phase boundary between the $s-$wave and the hybrid superconductor}
\label{app:phaseboundary}

To further corroborate the possibility of phase coexistence close to the line given by Eq.~\eqref{eq:phaseboundary-swave} [see also Fig.~\ref{fig:phasediagram}(a)], defined by  $\theta_0 = \pi/2$ ($\Delta_s=\bar{\Delta}_0$), we expand the expressions in Eq.~(\ref{eq:J3}) up to third order with respect to the small parameter $x = \frac{\pi}{2} - \theta_0 \ll 1$, and substitute them into Eq.~(\ref{eq:monop1}), to obtain
\bea
\lambda_s^{-1} &=& 2\ln(2\omega_D) - \ln(\Delta_s^2) - \frac{2}{3}x^3\nonumber\\
\lambda_{inter}^{-1} &=& \frac{4}{3}\ln(2\omega_D) -
\frac{2}{3}\ln(\Delta_s^2) - \frac{2}{3}x^2 + \frac{4}{3}x^3
\eea
We notice that for $x\ll 1$,
\bea
\lambda_{inter}^{-1} - \frac{2}{3}\lambda_{s}^{-1} = -\frac{2}{3}x^2
+ \frac{16}{9}x^3 < 0,
\eea
and hence the phase coexistence is possible. However,  the positive cubic term implies that eventually, for a sufficiently strong inter-FS pairing, the monopole SC phase dominates, as  explicitly found in the analysis presented around Eq.~\eqref{eq:phaseboundary-swave} in the main text.


\section{Impurity Scattering}
\label{app:impurityscattering}

\begin{figure*}[t!]
    \centering
    \includegraphics[scale=0.4]{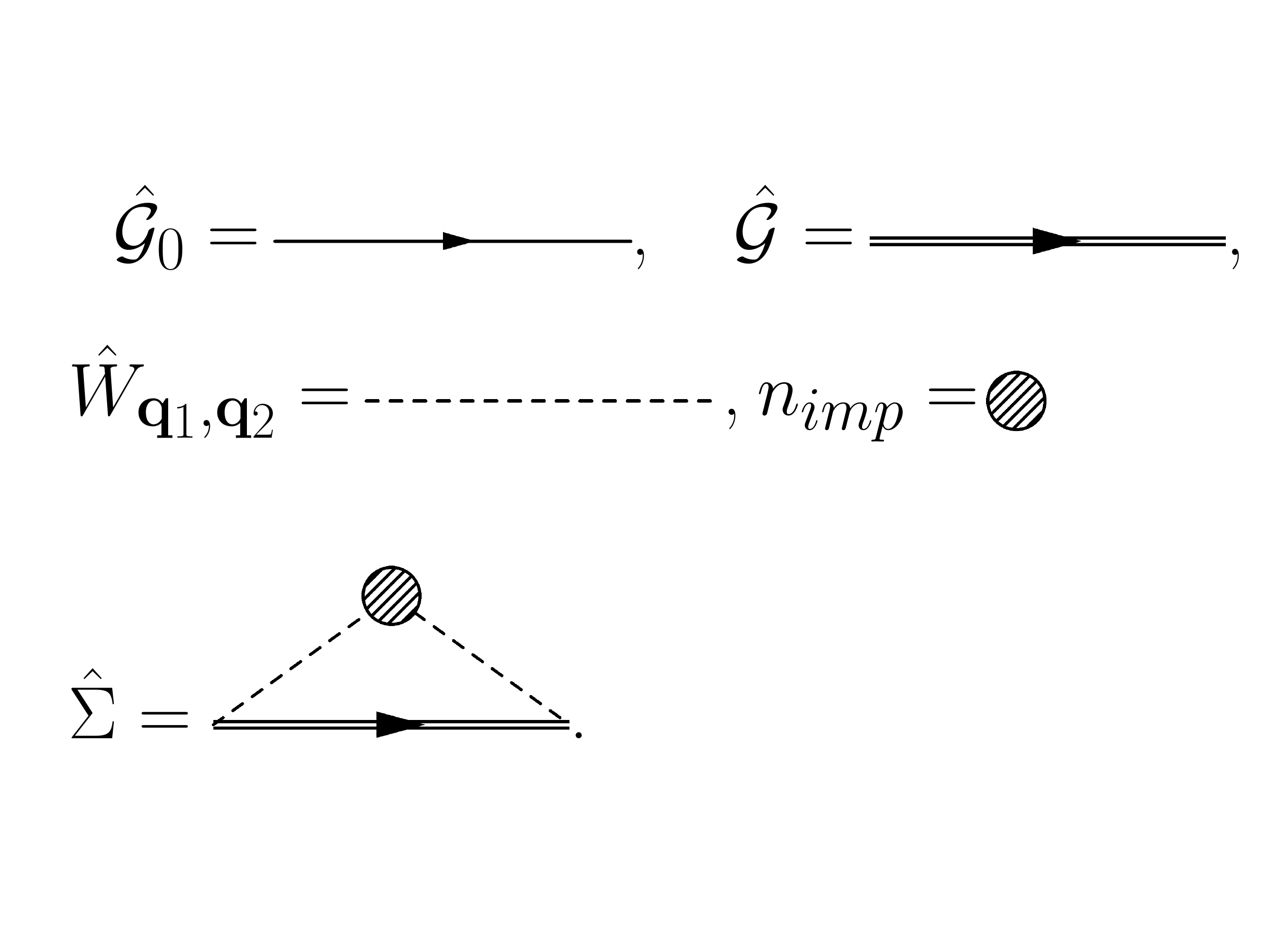}
    \caption{Feynman diagram for the impurity-averaged self-energy in the first Born approximation. The different symbols are also displayed.}
    \label{fig:Diag_1}
\end{figure*}

\begin{figure*}[t!]
    \includegraphics[scale=0.4]{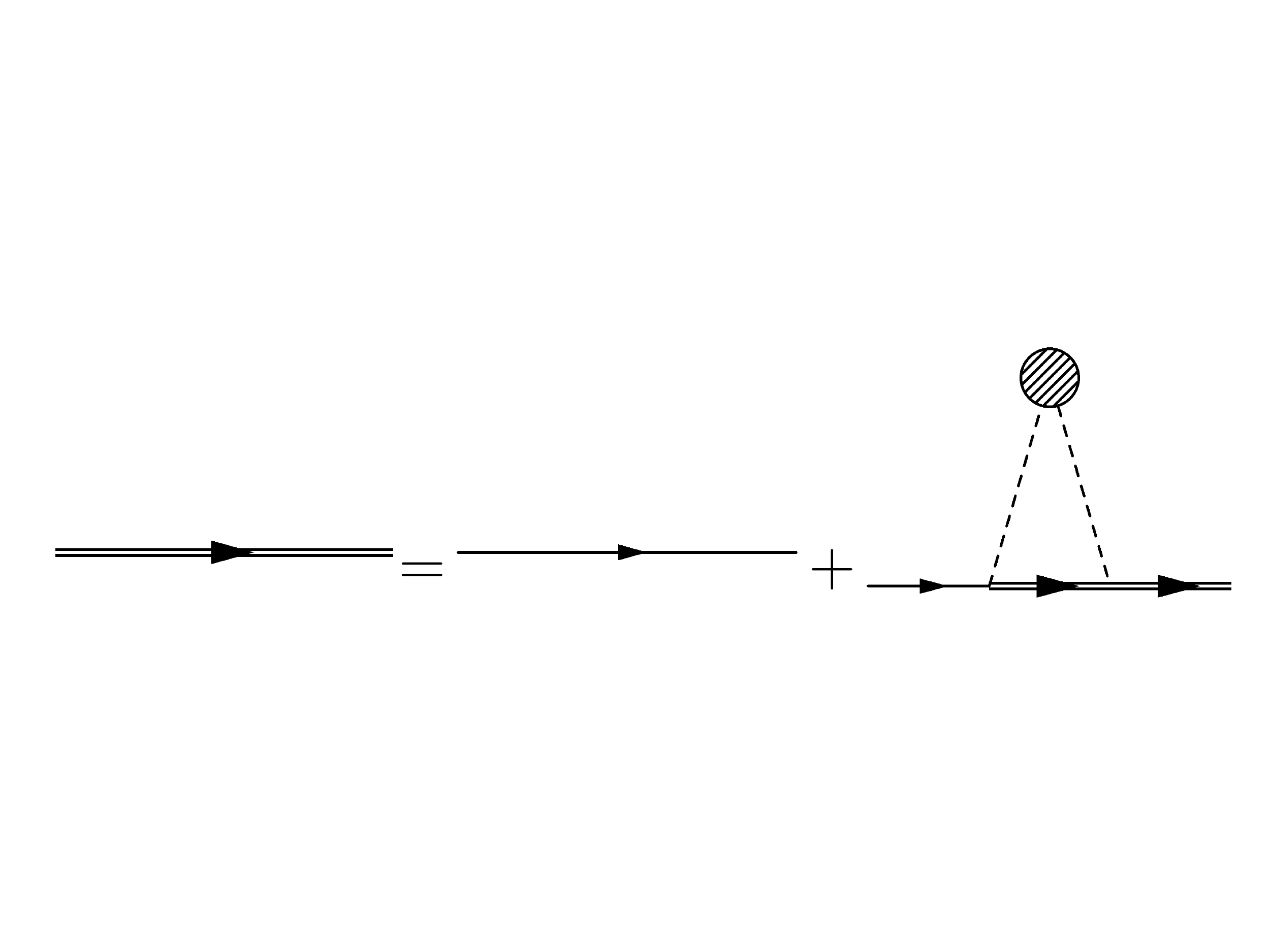}
    \caption{The Dyson equation for the first Born approximation. The symbols are described in Fig.~\ref{fig:Diag_1}}.
    \label{fig:Diag_2}
\end{figure*}

We now address the effect of scattering by randomly distributed, non-magnetic impurities, with a concentration $n_{imp}$. Within the first Born approximation, this can be captured through an averaged self-energy matrix, of the form \cite{Rammer-86}
\begin{equation}
\hat{\Sigma}(\omega_n,\mathbf{q}) = n_{imp}\sum_{\mathbf{q}_1}
\hat{W}_{\mathbf{q},\mathbf{q}_1}\hat{\mathcal{G}}(\omega_n,\mathbf{q}_1)\hat{W}_{\mathbf{q}_1,\mathbf{q}}
\label{eq_selfenergy}
\end{equation}
where $\hat{\mathcal{G}}(\omega_n,\mathbf{q})$ is the full interacting Green's function matrix, arising from the solution to the Dyson equation
\begin{equation}
\hat{\mathcal{G}}^{-1}(\omega_n,\mathbf{q}) = \hat{\mathcal{G}}^{-1}_0(\omega_n,\mathbf{q}) - \hat{\Sigma}(\omega_n,\mathbf{q})
\label{eq_Dyson}.
\end{equation}
The Feynman diagrams representing the bare and dressed Green's functions, as well as the self-energy within the first Born approximation \cite{Rammer-86} are depicted in
Fig.~\ref{fig:Diag_1}.
The self-consistent solution for the Dyson equation~(\ref{eq_Dyson}) can be represented diagrammatically as shown in Fig.~\ref{fig:Diag_2}.

We assume, for simplicity, that the elements of the scattering potential matrix are independent of momenta, and include both inter- and intra-FS scattering processes
\begin{eqnarray}
\hat{W}_{\mathbf{q},\mathbf{q}_1} = u\hat{\tau}_1\otimes\hat{\eta}_0 + v\hat{\tau}_0\otimes\hat{\eta}_3.
\label{eq_W}
\end{eqnarray}
The Dyson equation can be solved by assuming the ansatz that the fully dressed Green's function possesses the same structure of
the the one in the clean case, but with renormalized parameters $\omega_{n,R}$, $\Delta_{R}(\mathbf{q})$, $\Delta_{0,R}$ and $\xi_{n,q,R}$.
In the following, we consider the Eq.~(\ref{eq:Gomat}) for $\hat{\mathcal{G}}_0(\omega_n,\mathbf{q})$, neglecting the small nodal asymmetry in the chemical potentials $\delta\mu = 0$. We repeat a similar matrix analysis as in Appendices B and C, and solve the Dyson equation
to obtain the renormalized parameters in the form
\begin{widetext}
\begin{eqnarray}
i\omega_{n,R} &=& i \omega_n -  n_{imp}\int\frac{d^3 k}{(2\pi)^3}\left\{
-(u^2 + v^2)\frac{i\omega_n E_k^2}{E_k^4 - 4 B_k^2}
+ 2 u v \frac{B_k\xi_k}{E_k^4 - 4 B_k^2}
\right\},\nonumber\\
\xi_{n,q,R} &=& \xi_q + n_{imp}\int\frac{d^3 k}{(2\pi)^3}\left\{
-(u^2 + v^2)\frac{\xi_k E_k^2}{E_k^4 - 4 B_k^2}
+ 2 u v \frac{B_k i\omega_n}{E_k^4 - 4 B_k^2}\right\},\nonumber\\
\Delta_{0,R} &=& \Delta_0 +
n_{imp}\int\frac{d^3 k}{(2\pi)^3} (u^2 - v^2)\frac{2 B_k\tilde{\Delta}(\mathbf{k}) - E_k^2\Delta_0}{E_k^4 - 4 B_k^2}.
\end{eqnarray}
\end{widetext}
Let us focus first on the intra-FS scattering case, by setting $u = 0$. At the critical temperature $\bar{T}_c$,
we apply the condition $\Delta_0 \rightarrow 0$, such that in this limit the parameters $E_k \rightarrow \xi_k$, and $B_k \rightarrow 0$. Therefore, the linearized form for the renormalized Matsubara frequencies reduces to
\begin{eqnarray}
\omega_{n,R} &=& \omega_n + \omega_n n_{imp} v^2 \int\frac{d^3 k}{(2\pi)^3}\frac{\xi_k^2 + \omega_n^2}{\left( \xi_k^2 + \omega_n^2\right)^2}\nonumber\\
&=& \omega_n + n_{imp}v^2\omega_n\int_{-\omega_D}^{\omega_D}d\xi\frac{\rho(\xi)}{\omega_n^2 + \xi^2}\nonumber\\
&=& \omega_n + 2 n_{imp} v^2 \rho(\mu) \arctan\left( \omega_D/\omega_n\right)\nonumber\\
&=& \omega_n\left(1 + \frac{1}{2\tau_{intra}|\omega_n|}  \right)
\end{eqnarray}
where in the last step, we assumed $\omega_D\gg T_c$, such that $\arctan\left( \omega_D/\omega_n\right) \rightarrow (\pi/2){\rm{sgn}}(\omega_n)$. We have also defined the
intra-FS scattering relaxation time
\begin{eqnarray}
\tau_{intra}^{-1} = 2\pi n_{imp} v^2 \rho(\mu)
\end{eqnarray}
Similarly, the equation for the renormalized band dispersion reduces to
\begin{eqnarray}
\xi_{n,q,R} &=& \xi_q - v^2 n_{imp}\int_{-\omega_D}^{\omega_D}d\xi\frac{\xi\rho(\xi)}{\omega_n^2 + \xi^2}\nonumber\\
&=& \xi_q,
\end{eqnarray}
since the integrand is odd and hence the correction to the band structure vanishes.
Finally, the renormalized pairing reduces to
\begin{eqnarray}
\Delta_{0,R} &=& \Delta_0 + 2 n_{imp}v^2 \Delta_0 \rho(\mu)\int_{0}^{\omega_D}\frac{d\xi}{\omega_n^2 + \xi^2}\nonumber\\
&=& \Delta_0\left( 1 + \frac{1}{2\tau_{intra}|\omega_n|} \right)
\end{eqnarray}
If one considers now the effect of inter-FS scattering, i.e. by setting $u\ne 0$, $v\ne 0$, the same analysis as above leads to
\begin{eqnarray}
\Delta_{0,R} &=& \Delta_0\left( 1 + \frac{1}{2}\left(\tau_{intra}^{-1} - \tau_{inter}^{-1} \right)|\omega_n|^{-1} \right)\nonumber\\
\omega_{n,R} &=& \omega_n \left( 1 + \frac{1}{2}\left(\tau_{intra}^{-1} + \tau_{inter}^{-1} \right)|\omega_n|^{-1} \right)
\end{eqnarray}
where we defined
\begin{eqnarray}
\tau_{inter}^{-1} = 2\pi n_{imp} u^2 \rho(\mu)
\end{eqnarray}

\subsection{BCS gap equation}

The BCS gap equations that include the effect of scattering by random impurities are given by
\begin{widetext}
\begin{eqnarray}
\Delta_0 &=& V_0 \sum_{n}\int\frac{d^3q}{(2\pi)^3} \frac{\Delta_{0,R}}{\omega_{n,R}^2 + \xi_{q,R}^2}
= V_0 \sum_{n}\int\frac{d^3q}{(2\pi)^3} \frac{\Delta_0\left( 1 + \frac{1}{2}\left(\tau_{intra}^{-1} - \tau_{inter}^{-1} \right)|\omega_n|^{-1} \right)}{\omega_n^2\left( 1 + \frac{1}{2}\left(\tau_{intra}^{-1} + \tau_{inter}^{-1} \right)|\omega_n|^{-1} \right)^2 + \xi_q^2}\nonumber\\
&=& 2 V_0\rho(\mu) \sum_{n}\int_{0}^{\omega_D} d\xi \frac{\Delta_0\left( 1 + \frac{1}{2}\left(\tau_{intra}^{-1} - \tau_{inter}^{-1} \right)|\omega_n|^{-1} \right)}{\omega_n^2\left( 1 + \frac{1}{2}\left(\tau_{intra}^{-1} + \tau_{inter}^{-1} \right)|\omega_n|^{-1} \right)^2 + \xi^2}\nonumber\\
&=& 2 V_0 \rho(\mu)\sum_{n}\frac{\Delta_0\left( 1 + \frac{1}{2}\left(\tau_{intra}^{-1} - \tau_{inter}^{-1} \right)|\omega_n|^{-1} \right)}{|\omega_n|\left( 1 + \frac{1}{2}\left(\tau_{intra}^{-1} + \tau_{inter}^{-1} \right)|\omega_n|^{-1} \right)}\arctan\left(\frac{\omega_D}{|\omega_n|\left( 1 + \frac{1}{2}\left(\tau_{intra}^{-1} + \tau_{inter}^{-1} \right)|\omega_n|^{-1} \right)} \right).
\end{eqnarray}
Assuming, as before, that $\omega_D \gg T_c$, the previous expression simplifies to
\begin{eqnarray}
1 = V_0 \rho(\mu)\pi\sum_{n}\frac{1}
{|\omega_n| + \tau_{inter}^{-1}}= \frac{V_0 \rho(\mu)}{T_c}\left[ \ln\left(\frac{\Gamma_c}{2\pi T_c} \right)
- \psi\left(\frac{1}{2} + \frac{1}{2\pi\tau_{inter} T_c}  \right)
\right],
\label{eq:BCS-impurity1}
\end{eqnarray}
where $\Gamma_c$ is an upper cutoff for the Matsubara frequency sum, and $\psi(x)$ is the digamma function.
\end{widetext}

A similar analysis can now be performed for the monopole SC pairing. In this case, the vortex part protects the  gap function from renormalization,
\begin{eqnarray}
\tilde{\Delta}_{R}(\mathbf{q}) = \tilde{\Delta}(\mathbf{q}).
\end{eqnarray}
Therefore, the corresponding BCS gap equation for the monopole SC pairing becomes
\begin{align}
&\tilde{\Delta}_0 = \kappa \sum_{n}
\int \frac{d^3q}{(2\pi)^3}\frac{\tilde{\Delta}_0}{\bar{\omega}_n^2 + \xi_q^2}\\
&=  2 \kappa \rho(\mu)\sum_{n}
\int_\xi\frac{\tilde{\Delta}_0}{\omega_n^2\left(1 + \frac{1}{2}\left(\tau_{inter}^{-1} + \tau_{intra}^{-1} \right)|\omega_n|^{-1} \right)^2 + \xi^2},
\end{align}
where $\int_\xi\equiv \int_0^{\omega_D} d\xi$ and $\kappa$ is defined as in Eq.(\ref{eq:kappa}).
Analogously to the previous case, i.e. $\omega_D\gg T_c$, the integral and Matsubara sum can be performed to yield the corresponding equation for the critical temperature of the monopole pairing
\begin{equation}
\frac{T_c}{\kappa \rho(\mu)} = \ln\left(\frac{\Gamma_c}{2\pi T_c} \right)
- \psi\left( \frac{1}{2} + \frac{\tau_{intra}^{-1} + \tau_{inter}^{-1}}{4\pi T_c} \right).
\label{eq:BCS-impurity2}
\end{equation}
 Figure 3 in the main text is obtained by using Eqs.~\eqref{eq:BCS-impurity1} and~\eqref{eq:BCS-impurity2} and shows the dependence of the phase boundary on the intra-FS inverse scattering  time for various choices of $\tau_{inter}^{-1}$. At the phase boundary the critical temperatures for the two pairings are equal.

\end{document}